\documentclass[pdflatex,sn-mathphys-num]{sn-jnl}

\usepackage{graphicx}%
\usepackage{multirow}%
\usepackage{amsmath,amssymb,amsfonts}%
\usepackage{amsthm}%
\usepackage{mathrsfs}%
\usepackage[title]{appendix}%
\usepackage{xcolor}%
\usepackage{textcomp}%
\usepackage{manyfoot}%
\usepackage{booktabs}%
\usepackage{algorithm}%
\usepackage{algorithmicx}%
\usepackage{algpseudocode}%
\usepackage{listings}%
\usepackage{subfigure}
\usepackage{gensymb}
\usepackage{bm}
\usepackage{upgreek}

\newcommand{\ii}[0]{\rm{i}}
\newcommand{\mm}[0]{\rm{m}}

\graphicspath{{./figures/}}

\begin{document}

\title{Quantifying the effects of particle clustering in random thermoelastic composites -- numerical and mean-field analyses}

\author[1]{\fnm{Pawe{\l}} \sur{Ho{\l}obut}}\email{pholob@ippt.pan.pl}

\author[2]{\fnm{Micha{\l}} \sur{Majewski}}\email{mmajew@pjwstk.edu.pl}

\author*[1]{\fnm{Katarzyna} \sur{Kowalczyk-Gajewska}}\email{kkowalcz@ippt.pan.pl}

\affil*[1]{\orgname{Institute of Fundamental Technological Research, Polish Academy of Sciences}, \orgaddress{\street{Pawinskiego 5B}, \city{Warsaw}, \postcode{02-106}, \country{Poland}}}

\affil[2]{\orgname{Polish-Japanese Academy of Information Technology}, \orgaddress{\street{Koszykowa 86}, \city{Warsaw}, \postcode{02-008}, \country{Poland}}}

\abstract{The effect of space distribution of randomly-placed particles in a representative composite volume on the thermoelastic effective properties and local stress and strain distribution is analyzed. Quantitative assessment is performed using both the full-field finite element analyses and the mean-field interaction model, known also as a ``cluster'' model. The latter model is developed in the multi-family setting enabling one to study the mean stress and strain separately for each inclusion of the representative unit cell. The particles are assumed to be spherical and of equal size, while considered examples differ by the volume fraction of inclusions and mean nearest-neighbour distances.}


\maketitle

\section{Introduction}


 Nowadays, novel advanced materials are created not only by modification of material composition, e.g. by new alloying additions in metal alloys, but also by tailoring the material microstructure or nanostructure by optimized processing routes \cite{Fullwood10}. {Micromechanics} provides a powerful tool for the effective design of such processes as it enables one to estimate the macroscopic effective properties of heterogeneous materials knowing the local constitutive laws for the phases at the micro-level and morphological features of components within the representative volume. The key role in the micromechanical modelling is played by {a micro-macro transition scheme (averaging rule)} that makes it possible to estimate the sensitivity of the overall response to the changes in microstructure and local properties \cite{KachanovSevostianov18}.
	
With respect to their microstructure, heterogeneous materials can be subdivided into those with periodic substructure and those with random but statistically homogeneous one. In the first case, a unit cell for the microstructure can be defined which by its replication in three directions fills the material volume. Materials with this type of microstructure are mainly synthetic ones like e.g. meta-materials produced by additive manufacturing. In the second case, the substructure can be characterized by a set of statistical distribution functions of some microstructrural parameters, e.g. size and shape distributions or various n-point correlation functions \cite{Torquato02} such as the mean minimum distance between the particles. Development and validation of a mean-field model of particulate composites with random particle distribution will be the subject of the present paper.

While the use of numerical homogenization, in which the actual microstructure of a material can be directly subjected to mechanical analysis \cite{Segurado03,Segurado06,Ogierman25}, seems to be the obvious choice to study the impact of microstructural features on the overall composite response, such an approach is still relatively time-consuming. The use of analytical models allows testing many scenarios and better understanding of the microstructure-property relationship. 
The majority of existing analytical mean-field models are based on the Eshelby solution. The solution applies to a \emph{single ellipsoidal inclusion} in an infinite linearly elastic medium, subjected to a homogeneous external load at infinity. The Mori-Tanaka method based on this solution is the most widely used for composite materials. The model assumes that a single inclusion, being the representative of all the inhomogeneities in the representative volume, interacts only with the surrounding matrix while direct interactions between inclusions are neglected, which limits the applicability of the model to relatively small volume fractions. The other possibilities are the self-consistent and generalized self-consistent models, or differential and incremental schemes, which improve the predictions for higher volume contents or increased contrast between the phase properties, mainly by modifying the infinite matrix properties when applying the Eshelby result, cf.\cite{KachanovSevostianov18,Kachanov24}. 

Unfortunately, the above-mentioned classical models fail to describe properly the effects of spatial distribution of inhomogeneities within the sample, which may be an important source of a material's anisotropy, even if it is composed of isotropic phases, as well as they neglect the variability of mean strains and stresses within the individual inclusions. The issue of microstructure-induced anisotropy was studied in more detail in the previous paper by the authors \cite{Bieniek24}.  Let us only mention that \cite{Vilchevskaya21} distinguished two classes of approaches to dealing with the anisotropic effect of regular particle distribution: i. homogenization models based on the one-particle approximation, e.g.: \cite{PonteCastaneda95,HuWeng00,Sevostianov14,Sevostianov19}; and ii. analytical or semi-analytical solutions for periodic arrangements of identical inhomogeneities, e.g.:  \cite{Kushch97,Cohen04,schj05,Kushch11}. The packing effect (but not the spatial distribution effect) can also be accounted for by the so-called morphologically representative pattern (MRP) approach developed in \cite{Bornert96,Marcadon07} and verified through comparison with numerical homogenization in \cite{Majewski17} and \cite{Majewski20} for elastic and elastic-plastic composites, respectively.   An extensive review of the mentioned approaches can be found in \cite{Kowalczyk21,Bieniek24} or \cite{Vilchevskaya21}.

In \cite{Bieniek24}, when considering elastic-plastic composites with regular particle distribution, we selected and developed a different approach, originally proposed in \cite{MolinariElMouden96} -- \textbf{the cluster interaction model}, which includes the effect of morphological and spatial distribution of particles through the approximate solution of the Lippman-Schwinger-Dyson equation and the analytical result for \emph{two interacting inclusions} in the infinite medium by \cite{Berveiller87}. The model was later extended to coated inclusions \cite{ElMoudenCherkaoui98} and thermoelasticity  \cite{ElMoudenMolinari00}. The approach was also adopted to derive the thermal conductivity of a composite with coated inclusions in \cite{mercier00}. Recently, an extension of the approach to viscoplastic and elastic-viscoplastic composites was formulated by combining the scheme with the tangent additive law \cite{Kowalczyk21}.  An attempt at experimental validation of the effect was presented in \cite{Kowalczyk24} where samples with regular pores were produced using a 3D printing technique and tested in the regime of small strain to assess the anisotropy of elastic stiffness. In all contributions above, the validation and computational examples concern the single family variant of the method.

Following the problem review above, the goal of this research is two-fold: (i) development of efficient computational procedures for the multi-family variant of the cluster model and its broad verification, and (ii) analysis of the effect of packing on the overall thermoelastic properties and local response of particulate composites.  To our best knowledge, a mean-field model for thermoelastic metal matrix composites with comparable predictive capabilities, as concerns the response of individual inclusions in the representative volume, is not available in the existing literature. The proposed approach is foreseen to facilitate the study of damage development in composite materials, enabling assessment of different thresholds of damage initiation depending on the level of particle clustering \cite{QiWu19,NafarDastgerdi18}. Such variation was observed employing numerical homogenization techniques \cite{Segurado06,Weglewski13,Sequeira26}. These studies demonstrated that while clustering has minimal influence on the overall elastic properties, it leads to the increased spread of mean fields between particles. Furthermore, they showed that the damage growth rate of a composite with a higher degree of clustering is significantly faster than for composites with a lower degree of clustering. It can be noted here that \cite{ma04} also incorporated direct interactions between inclusions in the study of elastic-plastic composites, however, without introducing the concept of the cluster. The authors considered a random inclusion arrangement in the representative volume without its periodic extension to fill the whole space.

The paper consists of five sections. After the present introductory part, the formulation of the cluster interaction scheme in the context of the thermoelastic constitutive law is provided in Section~\ref{Sec:ClMod}. The attention is focused on the multi-family  variant of the method relevant for random particulate composites. Section~\ref{Sec:Numer} discusses details of the numerical finite element calculations performed to validate the approach. In Section~\ref{Sec:Results}, the variability of effective thermoelastic properties induced by specific arrangements of particles and the validity of results obtained by the analytical model are analyzed. In particular, the effect of the mean minimum distance describing the particle packing is quantified. Two aluminum alloy composite materials developed in \cite{Sequeira25} as well as a porous material are studied. Moreover, the variability of local mechanical fields in the particles is analyzed. The paper closes with conclusions. Additionally, two appendices provide the details of the calculation algorithm for the multi-family  variant of the cluster model and discuss the necessary cluster size to achieve convergent results.

The following notation is used for vector and tensor operations: the full contraction of the second-order tensor $\mathbf{a}$ with the fourth-order tensor $\mathbb{A}$, resulting in another second-order tensor $\mathbf{b}$, is denoted by the '$\,\cdot\,$' symbol: $\mathbf{b}=\mathbb{A}\cdot\mathbf{a}$ or $\mathbf{b}=\mathbf{a}\cdot\mathbb{A}$ meaning $b_{ij}=A_{ijkl} a_{kl}$ or $b_{ij}=a_{kl}A_{klij}$ in the components of the quantities in any orthonormal basis; the partial contraction of two fourth-order tensors $\mathbb{A}$ and $\mathbb{B}$ resulting in another fourth-order tensor $\mathbb{C}$ is denoted as: $\mathbb{C}=\mathbb{A}\mathbb{B}$ meaning $C_{ijkl}=A_{ijmn}B_{mnkl}$. Summation convention over repeated indices is used when not stated otherwise.

\section{Mean-field interaction model for thermoelastic composites\label{Sec:ClMod}}

In the interaction model, proposed originally by \cite{MolinariElMouden96} for elastic composites, the composite microstructure is represented by an elementary cubic volume  (a representative unit cell RUC) containing a finite {number $N$ of particles}. When periodically reproduced, the RUC is filling the whole space. In this paper we assume a random spatial layout of spherical particles of equal sizes in the elementary volume, which is periodic at its boundaries. To each inclusion of the RUC we assign a set of inclusions which are its periodic images in the filled space. All inclusions in the set are supposed to be subjected to the same thermomechanical fields. This set of geometrically equivalent particles is called \emph{a family}. Contrary to regular arrangements studied in \cite{Bieniek24}, the inclusions within the RUC are not symmetrically equivalent and may exhibit different mean strains and stresses, so the number of families in this case is equal to the number $N$ of inclusions in the RUC. As a consequence, a multi-family variant of the cluster model is required in the present study. We concentrate our attention on the effect of particle clustering on the effective properties as well as stress and strain heterogeneity among different particles within the cell stemming from their location with respect to the neighboring particles.      

The mean-field interaction model (also known as a cluster model) was extended to the thermoelastic composite materials in \cite{ElMoudenMolinari00}.
The core idea of the model is to account for interactions between each pair of inclusions within the analyzed space, in addition to interactions between the inclusions and the matrix, by using the fundamental solution by Green's function technique (see \cite{Berveiller87}). Since, according to this solution, mutual interactions experienced by the selected inclusion in the RUC decay when the distance to the other inclusion increases, the analysis is restricted only to those inclusions which are contained in some material sub-volume of spherical shape. This spherical sub-domain is centered in the selected inclusion for which the influence of such interactions is assessed.  The radius of the interaction sub-domain is established by some preliminary studies. In the considered examples, the radius 10 times larger than the RUC size was found to be sufficient (see Appendix \ref{Sec:App2}).

\begin{figure}[ht]
\centering
\begin{tabular}{cc}
(a) & (b)\\
\includegraphics[width=0.45 \textwidth]{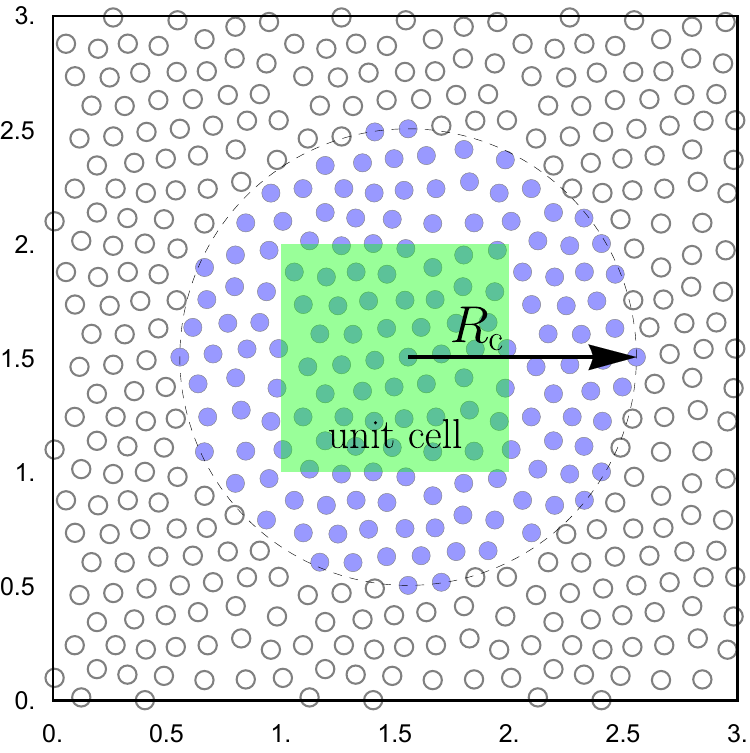}&
\includegraphics[width=0.45 \textwidth]{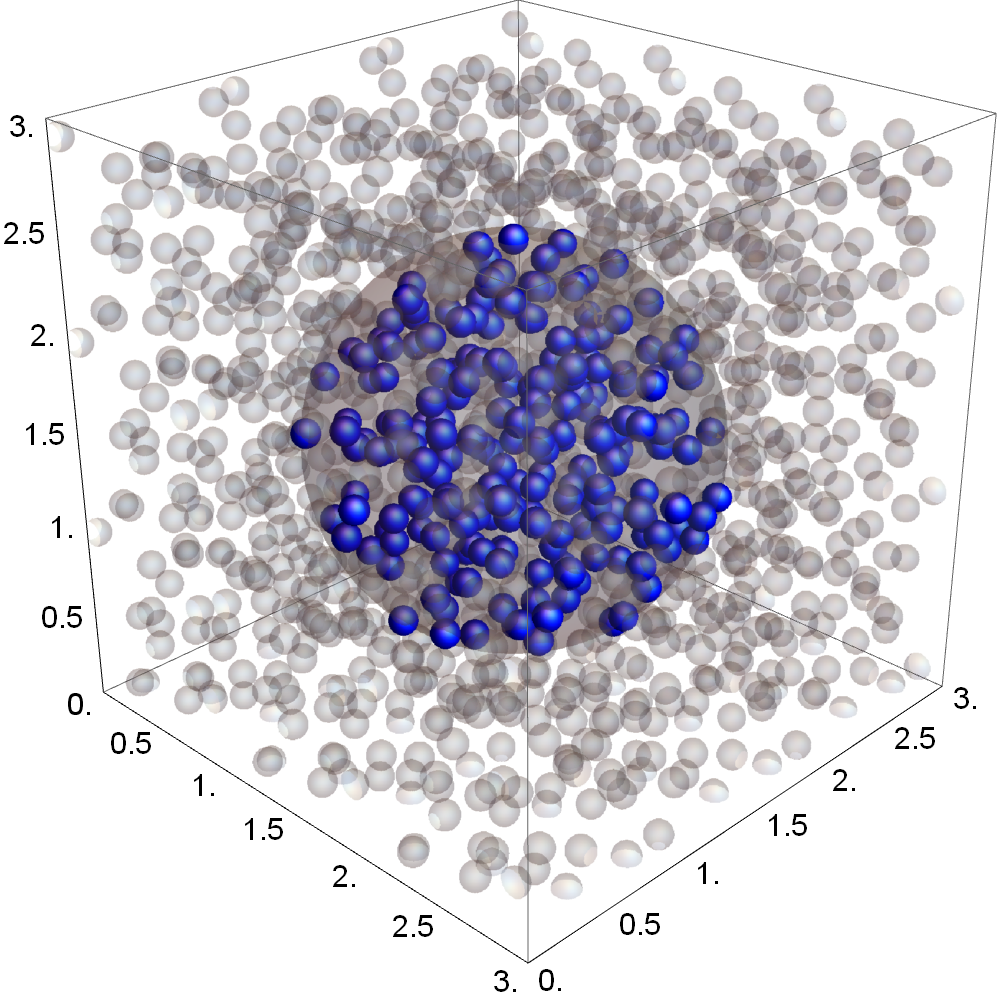}
\end{tabular}
\caption{
(a) {2D cluster composed of the inclusions (in blue) whose centers are on or inside the circle of radius $R_{\rm{c}}$.} 
(b) {3D cluster.}}.
\label{Fig:clusterradius}
\end{figure}

The basic relations of the model developed in \cite{ElMoudenMolinari00} are now recalled. The proposed efficient computational procedure is provided in the Appendix.    
The local constitutive relation between the stress $\boldsymbol{\sigma}$ and strain $\boldsymbol{\varepsilon}$ in each phase $r$ is formulated as follows 
\begin{equation}\label{Eq:tel}
\boldsymbol{\sigma}=\mathbb{C}_r\cdot\boldsymbol{\varepsilon}-\boldsymbol{\beta}_r\theta\,,
\end{equation}
where $\mathbb{C}_r$ is the fourth-order stiffness tensor for phase $r$, and $\boldsymbol{\beta}_r\theta$ is the thermal stress at temperature change $\theta$ uniform in the composite ($\boldsymbol{\beta}_r={\mathbb{C}_r}\cdot\boldsymbol{\alpha}_r$, and $\boldsymbol{\alpha}_r$ is the linear thermal expansion tensor). A two-phase composite is considered, so all inclusions in the RUC have the same material properties uniform per phase: $\mathbb{C}_{\ii}$, $\boldsymbol{\beta}_{\ii}$. The respective matrix properties are denoted by $\mathbb{C}_{\mm}$ and $\boldsymbol{\beta}_{\mm}$.
Similar relations are also valid for the respective mean values $\boldsymbol{\sigma}_I=\left<\boldsymbol{\sigma}\right>_I$ and $\boldsymbol{\varepsilon}_I=\left<\boldsymbol{\varepsilon}\right>_I$ over volume $V_I$ occupied by family $I$ ($I=1,\ldots,N$). $<.>_I=1/V_I\int_{V_I}(.)\,\rm{d}V$ denotes the operation of volume averaging over the inclusions belonging to the family $I$, as well as for the matrix, namely:
\begin{equation}\label{Eq:tel-av}
\boldsymbol{\sigma}_I=\mathbb{C}_{\ii}\cdot\boldsymbol{\varepsilon}_I-\boldsymbol{\beta}_{\ii}\theta\,,\quad 
\boldsymbol{\sigma}_{\mm}=\mathbb{C}_{\mm}\cdot\boldsymbol{\varepsilon}_{\mm}-\boldsymbol{\beta}_{\mm}\theta\,.
\end{equation}

As in any mean-field model,  the effective stiffness tensor $\bar{\mathbb{C}}$ and effective thermal stress $\bar{\boldsymbol{\beta}}\theta$ can be found for the composite representative volume $V$, such that
\begin{equation}\label{Eq:tot}
\bar{\boldsymbol{\sigma}}=\sum_{I=1}^Nf_I\boldsymbol{\sigma}_I+(1-f_{\ii})\boldsymbol{\sigma}_{\mm}=	\bar{\mathbb{C}}\cdot\bar{\boldsymbol{\varepsilon}}-\bar{\boldsymbol{\beta}}\theta\,,\quad
\bar{\boldsymbol{\varepsilon}}=\sum_{I=1}^Nf_I\boldsymbol{\varepsilon}_I+(1-f_{\ii})\boldsymbol{\varepsilon}_{\mm}\,,
\end{equation}
where $f_{\ii}$ is the volume fraction of all inclusions in the RUC and $\sum_I f_I=f_{\ii}$.
As the local constitutive law is affine with $\mathbb{C}_r$ and $\boldsymbol{\beta}_r$ uniform per phase,
 the effective properties are specified using the relevant mechanical strain localization tensors  $\mathbb{A}_I$ and $\mathbb{A}_{\mm}$ of the fourth order:
\begin{equation}\label{Eq:C}
\bar{\mathbb{C}}=\mathbb{C}_{\ii}\left({\sum_{I=1}^N} f_I\mathbb{A}_I\right)+(1-f_{\ii})\mathbb{C}_{\mm}\mathbb{A}_m\,,
\end{equation}
\begin{equation}\label{Eq:Beta}
\bar{\boldsymbol{\beta}}=\boldsymbol{\beta}_{\ii}\cdot\left({\sum_{I=1}^N} f_I\mathbb{A}_I\right)+(1-f_{\ii})\boldsymbol{\beta}_{\mm}\cdot\mathbb{A}_{\mm}\,.    
\end{equation}
Based on \eqref{Eq:C} and \eqref{Eq:Beta}, the mechanical strain localization tensor for the whole inclusion phase can be specified as:
\begin{equation}\label{Eq:Aimean}
\mathbb{A}_{\ii}= \frac{1}{f_{\ii}}{\sum_{I=1}^N} f_I\mathbb{A}_I   
\end{equation}
The mean strain in each inclusion in the RUC, as a mean strain for the given family, and the mean-strain in the matrix phase can be specified separately by the strain localization relations of a standard form 
\begin{equation}\label{Eq:loc}
\boldsymbol{\varepsilon}_I=\mathbb{A}_I\cdot\bar{\boldsymbol{\varepsilon}}+\mathbf{e}_I\theta\quad (I=1,\ldots,N)\,,\quad
\boldsymbol{\varepsilon}_{\mm}=\mathbb{A}_{\mm}\cdot\bar{\boldsymbol{\varepsilon}}+\mathbf{e}_{\mm}\theta\,.
\end{equation}
When the mechanical strain localization tensors $\mathbb{A}_I$, $\mathbb{A}_{\mm}$ and the thermal strain localization tensors of the second order: $\mathbf{e}_I$, $\mathbf{e}_{\mm}$  are known, the set of relations (\ref{Eq:tel}--\ref{Eq:loc}) enables one to find the mean stresses and strains per each family and the matrix, as well as macroscopic stress or strain, depending on which of these two quantities is prescribed by the boundary conditions.

Following \cite{ElMoudenMolinari00}, and notation introduced in \cite{Bieniek24}, it can be demonstrated that for the multi-family variant of the interaction model, the mechanical strain localization tensors for the families and the matrix are found from the following set of fourth-order equations:

\begin{equation}\label{Eq:set-thermoel}
\sum_{K=1}^{N}\mathbb{M}^{IK}\mathbb{A}_K=\mathbb{I},\quad I=1,2,\ldots,N
\end{equation} 
and 
\begin{equation}\label{Eq:Am}
\mathbb{A}_{\mm}=\frac{1}{1-f_{\ii}}\left(\mathbb{I}-f_{\ii}\mathbb{A}_{\ii}\right)
\end{equation} 
where $\mathbb{A}_{\ii}$ is specified by \eqref{Eq:Aimean}.
The fourth-order tensors $\mathbb{M}_{IK}$ are specified as
\begin{equation}\label{Eq:MIK}
\mathbb{M}^{IK}=\left\{\begin{array}{lcl}
\mathbb{I}+\left((1-f_{K})\mathbb{P}_0-\bar{\boldsymbol{\Gamma}}^{IK}\right)(\mathbb{C}_{\ii}-\mathbb{C}_{\mm})&\textrm{for}&K=I\\
-\left(\bar{\boldsymbol{\Gamma}}^{IK}+f_{K}\mathbb{P}_0 \right)(\mathbb{C}_{\ii}-\mathbb{C}_{\mm})&\textrm{for}&K\neq I
\end{array}
\right.
\end{equation} 
The tensor $\mathbb{P}_0$ is the so-called polarisation tensor for a spherical inclusion immersed in the matrix material (see Eq. \eqref{Eq:Po}). The tensors $\bar{\boldsymbol{\Gamma}}^{IK}$ are conveying averaged mutual interaction between inclusions belonging to two families $I$ and $K$. They are obtained by summing the interactions between each inclusion pair $I$ and $j$ such that $I$ denotes the reference inclusion of the family $I$ \emph{in the elementary cube (i.e. RUC)}, while inclusion $j$  belongs to the family $K$ and is one of $N_K$ inclusions which are obtained by periodic expansion of the elementary cube, namely:
\begin{equation}\label{Eq:AGamma}
\bar{\boldsymbol{\Gamma}}^{IK}=\left\{
\begin{array}{lcl}
\sum_{j=2}^{N_K}\boldsymbol{\Gamma}^{Ij}&\textrm{for}&K=I\\
\sum_{j=1}^{N_K}\boldsymbol{\Gamma}^{Ij}&\textrm{for}&K\neq I
\end{array}
\right.
\end{equation}

For isotropic properties of the matrix and the spherical shape of inclusions, tensors $\mathbf{\Gamma}^{Ij}$ are specified in the Appendix A of \cite{Bieniek24} (see also \cite{MolinariElMouden96}). Note that for the multi-family case, $a$ and $b$ in these formulas are now the radii of inclusions belonging to the family $I$ and $K$, respectively ($a=b$ in the present analysis of inclusions of equal sizes) and $R$ is a distance between the reference inclusion of the family $I$ \emph{in the elementary cube (i.e. RUC)} and the inclusion $j$ of family $K$ in the cluster. 

The second-order thermal strain localization tensors $\mathbf{e}_I$ ($I=1,2,\ldots,N$) for the thermoelastic problem are found from the related set of $N$ second-order tensor equations 
\begin{equation}\label{Eq:set-thermoel2}
\sum_{I=1}^{N}\mathbb{M}^{IK}\cdot\mathbf{e}_K=\bar{\mathbf{w}}^I,\quad K=1,2,\ldots,N
\end{equation}
and 
\begin{equation}\label{Eq:em}
\mathbf{e}_{\mm}=-\frac{1}{1-f_{\ii}}\sum_{I=1}^N f_I\mathbf{e}_I\,.
\end{equation}
The fourth-order tensors $\mathbb{M}^{IK}$ are specified above, while the second-order tensors $\bar{\mathbf{w}}^I$ are specified as:
\begin{equation}\label{Eq:wI}
\bar{\mathbf{w}}^I=\sum_{L=1}^{N}\mathbf{w}^{IL}\,\, \textrm{where}\,\,\mathbf{w}^{IL}=\left\{\begin{array}{lcl}
\left((1-f_L)\mathbb{P}_0-\bar{\boldsymbol{\Gamma}}^{IL}\right)\cdot(\boldsymbol{\beta}_{\ii}-\boldsymbol{\beta}_{\mm})&\textrm{for}&L=I\\
-\left(\bar{\boldsymbol{\Gamma}}^{IL}+f_L\mathbb{P}_0 \right)\cdot(\boldsymbol{\beta}_{\ii}-\boldsymbol{\beta}_{\mm})&\textrm{for}&L\neq I
\end{array}
\right.
\end{equation}
with $\bar{\boldsymbol{\Gamma}}^{IL}$ and $\mathbb{P}_0$ specified before.

Let us observe that the solution of the set of tensorial equations \eqref{Eq:set-thermoel} and \eqref{Eq:set-thermoel2} requires an extension of the classical procedures used to solve linear equations. The extended Gauss elimination procedure is provided in the Appendix \ref{Sec:App1}.

\section{Numerical full-field model \label{Sec:Numer}}

\begin{figure}
	\centering
	\subfigure{(a)}{\includegraphics[angle=0,width=0.45\textwidth]{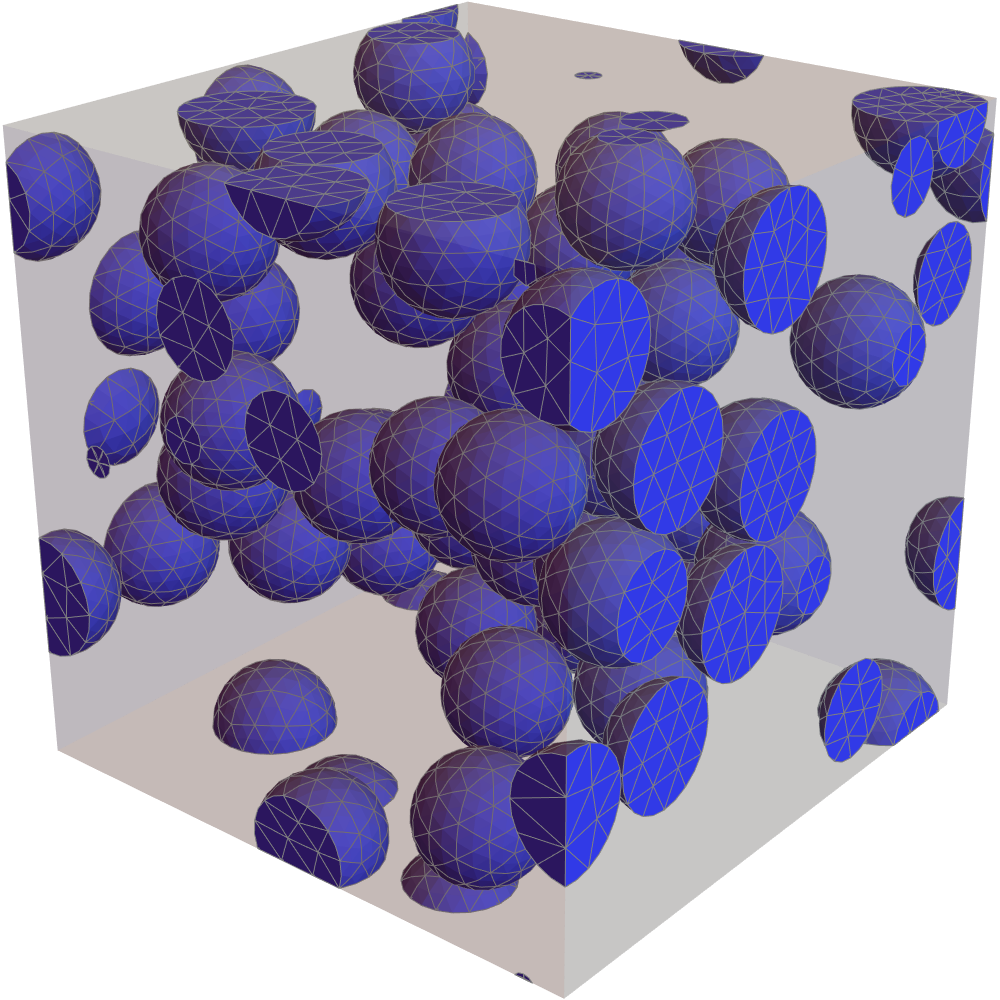}}
	\hfill
	\subfigure{(b)}{\includegraphics[angle=0,width=0.45\textwidth]{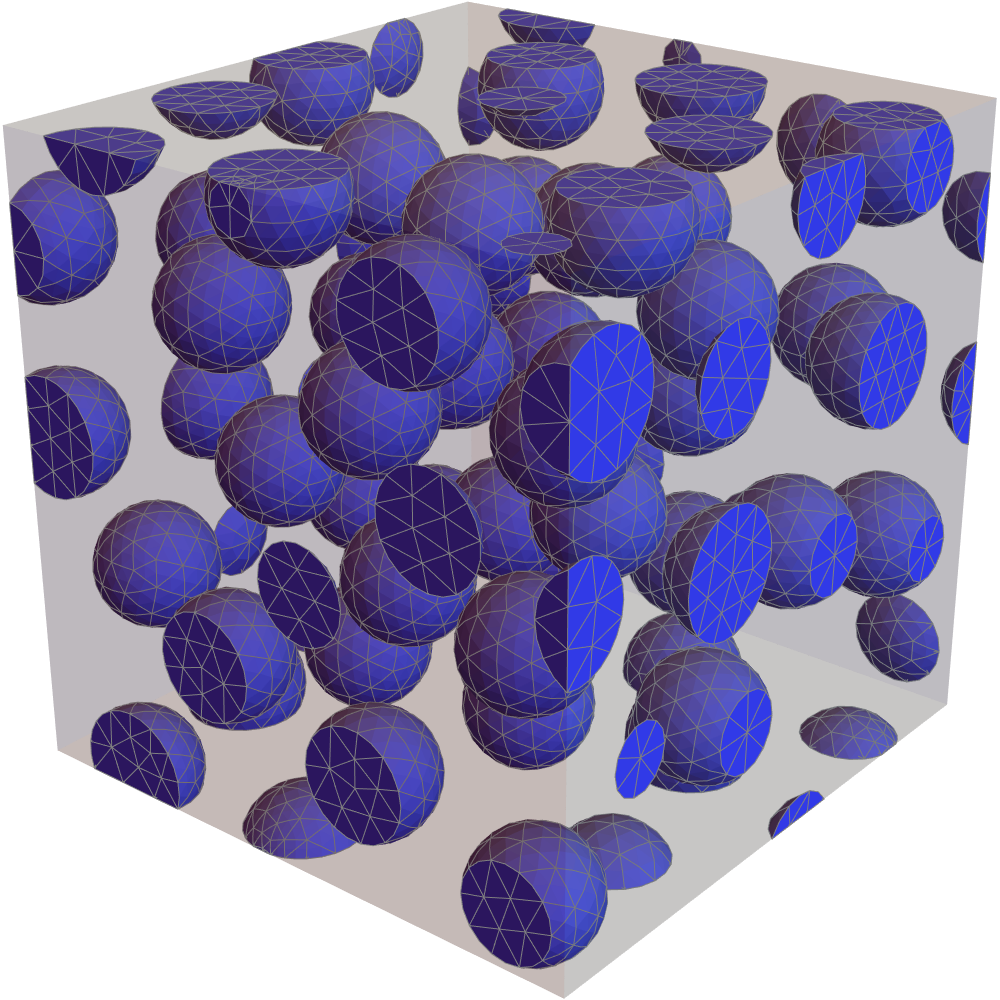}}
	\caption{Periodic RUCs with 50 randomly-placed inclusions occupying a volume fraction $f_{\ii}=0.2$, with the FE mesh shown only for the inclusions: (a) $\bar{\lambda}/(2R_{\ii}) = 0.008$, and (b) $\bar{\lambda}/(2R_{\ii}) = 0.454$.}
	\label{Fig:example-rucs-c02}
\end{figure}

Mean-field estimates obtained by the cluster model are compared below with the results of numerical homogenization performed using the Finite Element Method (FEM), for random microstructures with non-overlaping spherical particles of equal radius. The FE analyses were performed in the AceFEM environment \cite{korelc2002multi}, with the structural FE meshes for the  RUCs generated using NetGen \cite{schoberl1997netgen}. Thirty \emph{RUCs with 50 randomly-placed inclusions} were generated and analyzed for each of the volume fractions of inclusions $f_{\ii}$ equal to $0.1$, $0.2$ and $0.3$. RUCs within each group differed in their mean nearest-neighbour distances $\bar{\lambda}/(2R_{\ii})$, ranging from around 0.035 to 0.815 for $f_{\ii}=0.1$, from 0.008 to 0.454 for $f_{\rm{i}}=0.2$---corresponding to the two RUCs shown in Fig.~\ref{Fig:example-rucs-c02}, and from 0.004 to 0.280 for $f_{\ii}=0.3$.

All RUCs were generated in a $1\times 1\times 1$ cube. The cube was assumed to be a periodic cell, with the inclusions crossing the boundary copied periodically to the opposite sides of the cube. The random placement of inclusions was performed using Yade \cite{Yade}, a Discrete Element Method (DEM) system. The adopted approach followed the one used in \cite{Zielinski15} and exploited the dynamics of frictionless elastic spheres. To have greater control over the mean nearest-neighbour distance of the generated microstructures---in particular to facilitate obtaining values of $\bar{\lambda}/(2R_{\ii})$ near the lowest and highest ends of the spectrum---some modifications of these algorithms were introduced. They were described in detail in \cite{Majewski17}.

For the purpose of performing FEM analyses, the generated RUCs were meshed using tetrahedral, 10 node, second-order finite elements. The final mesh contained between 25 and 40 thousand elements, depending on the volume fraction of inclusions and the particular random geometry. The adopted mesh density guaranteed sufficient accuracy of results at a moderate computational cost, compared with more time-consuming computations based on finer meshes. Each of the geometrically-periodic samples was subjected to micro-periodic displacement boundary conditions to assure periodicity of deformation.

\section{Results\label{Sec:Results}}

Three types of FEM simulations were performed to validate the predictions of the cluster model. The first one aimed at determining effective elastic parameters of the composites. Six independent deformation tests were performed for each microstructure, with the global deformation tensor $\bar{\boldsymbol{\varepsilon}}$ of the RUC having just one nonzero component in each test, namely $\bar{\varepsilon}_{11}=\delta$ or $\bar{\varepsilon}_{22}=\delta$ or $\bar{\varepsilon}_{33}=\delta$ or $2\bar{\varepsilon}_{12}=\delta$ or $2\bar{\varepsilon}_{13}=\delta$ or $2\bar{\varepsilon}_{23}=\delta$, with $\delta=10^{-4}$ in all tests. From the average stresses in the RUC obtained in the six tests, the full effective stiffness tensor $\bar{\mathbb{C}}$ of the composite was determined in each case. This allowed further calculation of the values of the effective bulk and shear moduli of the composite, $\bar{K}$ and $\bar{G}$, respectively, computed for an isotropic material whose stiffness tensor was the closest to $\bar{\mathbb{C}}$ in terms of Euclidean distance. The second type of tests were thermal expansion simulations, aimed at determining the average thermal expansion coefficients $\bar{\alpha}$ for each composite. They consisted in assigning a static, uniform temperature increase $\Delta T=10\degree\rm{C}$ to the entire RUC and allowing its unconstrained expansion. The value of $\bar{\alpha}$ was determined from the overall volumetric strain $\bar{\varepsilon}_{kk}$ of the RUC. The third type of tests were uniaxial stress simulations, performed by assigning $\bar{\varepsilon}_{33}=\delta$ as the global deformation of a given RUC in the direction $x_3$, and allowing free deformation of the RUC in the remaining directions. In all tests, average strains and stresses in all inclusions and in the matrix were recorded.

Three types of composites were numerically investigated. The first one was made of $\rm{AlSi_{12}}$ matrix reinforced with $\rm{Al_2O_3}$ inclusions, the second of $\rm{AlSi_{12}}$ matrix reinforced with $\rm{SiC}$ inclusions, and the third one was $\rm{AlSi_{12}}$ matrix with voids. The values of the elastic thermomechanical parameters of the constituent materials are presented in Table~\ref{Tab:material-parameters}. The voids were simulated as weakened matrix material, having $10^{-4}$ the stiffness of $\rm{AlSi_{12}}$, instead of being modeled as actual empty space. This was to facilitate the gathering of results, particularly of void deformations, which could be gleaned in the present scheme from the deformation of the FE mesh filling the voids. 

\begin{table}[ht]
	\centering 
    {\small
	\begin{tabular}{cccc}
		\hline
		  Material & $E [\rm{GPa}]$ & $\nu$ &  $\alpha [10^{-6}/\degree\rm{C}]$
		\\
		\hline
		  $\rm{AlSi_{12}}$ & 70 & 0.35 & 23.7 \\
		  $\rm{Al_2O_3}$ & 380 & 0.22 & 6.5 \\
		  $\rm{SiC}$ & 410 & 0.15 & 4.0 \\
		\hline
	\end{tabular}
    }
	\caption{The material parameters of $\rm{AlSi_{12}}$, $\rm{Al_2O_3}$ and $\rm{SiC}$ \cite{Sequeira25} 
    used in simulations: the Young's modulus $E$, the Poisson's ratio $\nu$ and the thermal expansion coefficient $\alpha$.} 
	\label{Tab:material-parameters}
\end{table}

\begin{figure}
\centering
\subfigure{(a)}{\includegraphics[angle=0,width=0.45\textwidth]{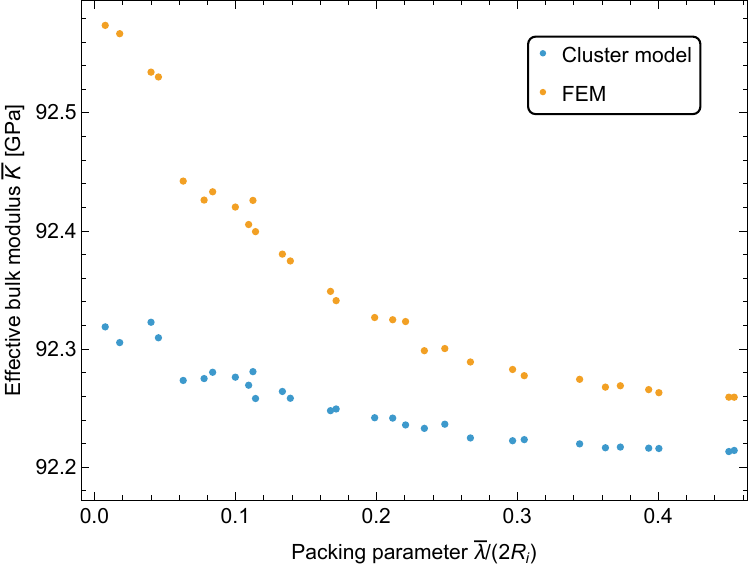}}
\hfill
\subfigure{(b)}{\includegraphics[angle=0,width=0.45\textwidth]{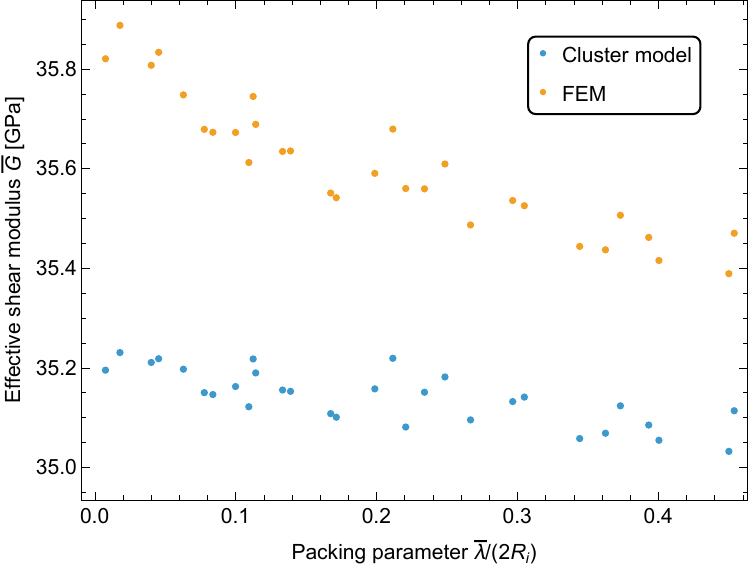}}
\caption{
Cluster model and FEM estimations of (a) the effective bulk modulus $\bar{K}$ and (b) the shear modulus $\bar{G}$ of the $\rm{AlSi_{12}}$ matrix reinforced with $\rm{Al_2O_3}$ particles (Tab.\ref{Tab:material-parameters}) vs. the packing parameter $\bar{\lambda}/(2R_{\ii})$. The RUC contains 50 inclusions occupying a volume fraction $f_{\ii}=20\%$. 
\label{Fig:KG20-Al2O3}}
\end{figure}

\begin{figure}
\centering
\subfigure{(a)}{\includegraphics[angle=0,width=0.45\textwidth]{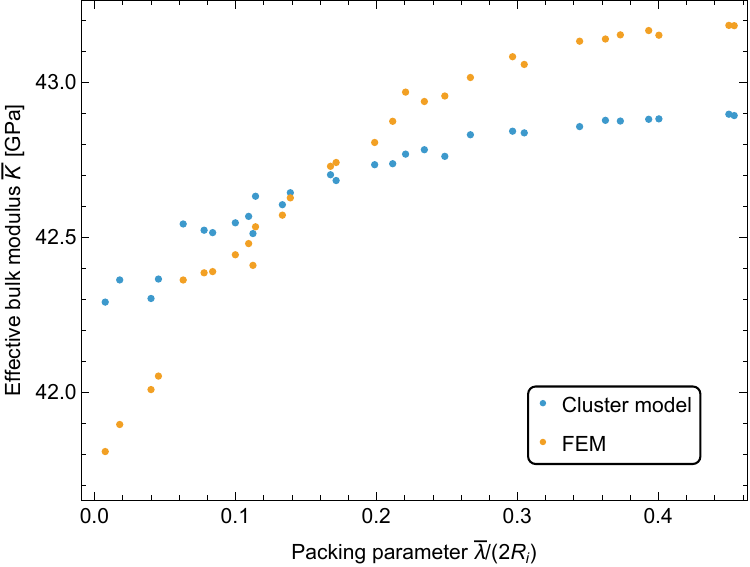}}
\hfill
\subfigure{(a)}{\includegraphics[angle=0,width=0.45\textwidth]{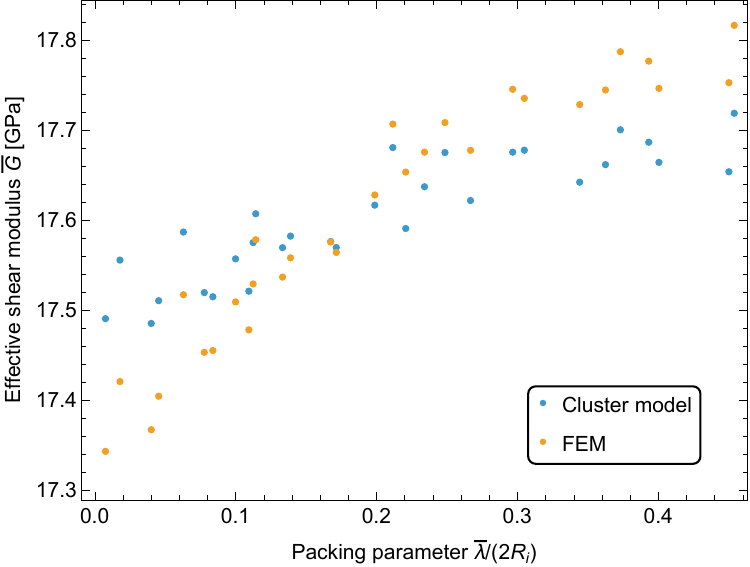}}
\caption{
Cluster model and FEM estimations of (a) the effective bulk modulus $\bar{K}$ and (b) the shear modulus $\bar{G}$ of the $\rm{AlSi_{12}}$ matrix with voids (Tab.\ref{Tab:material-parameters}) vs. the packing parameter $\bar{\lambda}/(2R_{\ii})$. The RUC contains 50 voids occupying a volume fraction $f_{\ii}=20\%$. 
\label{Fig:KG20-voids}}
\end{figure}

\begin{figure}
\centering
\subfigure{(a)}{\includegraphics[angle=0,width=0.45\textwidth]{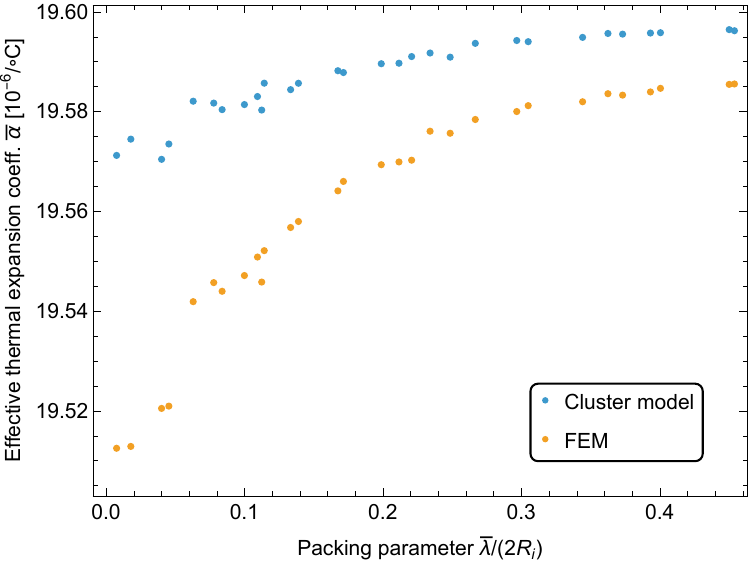}}
\hfill
\subfigure{(b)}{\includegraphics[angle=0,width=0.45\textwidth]{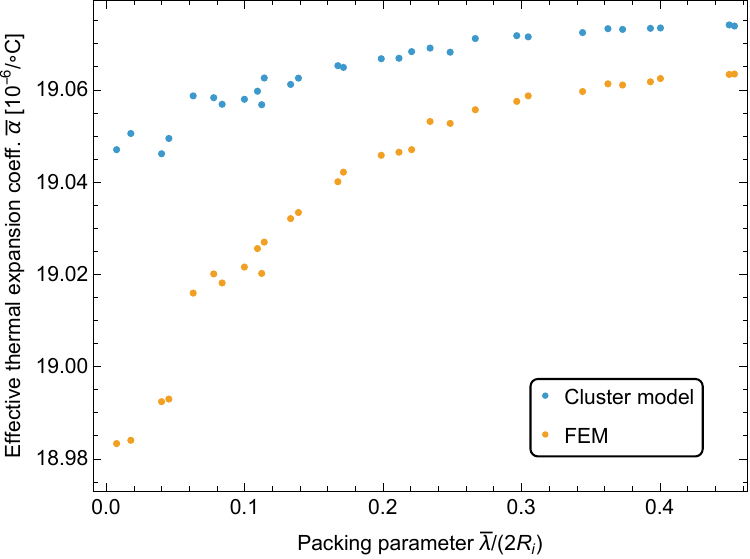}}
\caption{
Cluster model and FEM estimations of the effective thermal expansion coefficient $\bar{\alpha}$ of the $\rm{AlSi_{12}}$ matrix reinforced with (a) $\rm{Al_2O_3}$ and (b) $\rm{SiC}$ particles (Tab.\ref{Tab:material-parameters}) vs. the packing parameter $\bar{\lambda}/(2R_{\ii})$. The RUC contains 50 inclusions occupying a volume fraction $f_{\ii}=20\%$. 
\label{Fig:A20}}
\end{figure}

\begin{figure}
\centering
\subfigure{(a)}{\includegraphics[angle=0,width=0.45\textwidth]{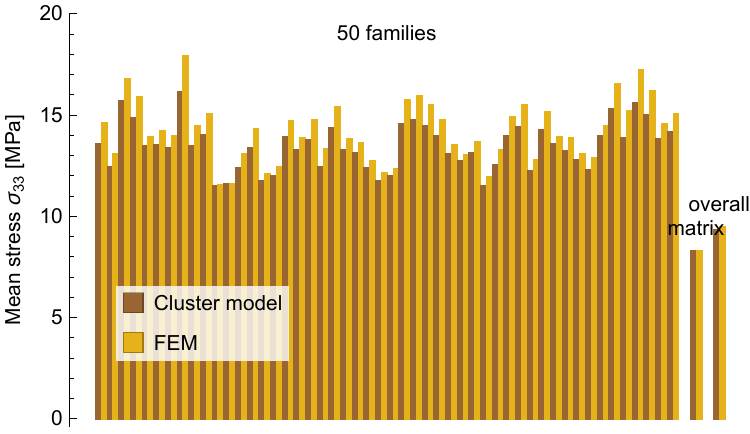}}
\hfill
\subfigure{(b)}{\includegraphics[angle=0,width=0.45\textwidth]{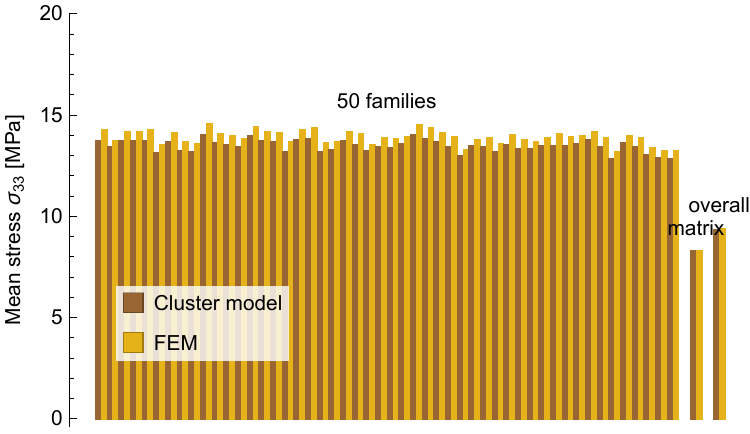}}
\caption{Mean stress $\sigma_{33}$ in all 50 $\rm{Al_2O_3}$ inclusions, in the $\rm{AlSi_{12}}$ matrix and in the entire RUC, obtained by the cluster model and FEM for the microstructures in Fig.~\ref{Fig:example-rucs-c02}, (a) and (b) respectively, under uniaxial stress of the RUC along $x_3$ at global strain $\bar{\varepsilon}_{33}=10^{-4}$.}
\label{Fig:n50_t0c02_Al2O3_bar_s33}
\end{figure}

\begin{figure}
\centering
\subfigure{(a)}{\includegraphics[angle=0,width=0.45\textwidth]{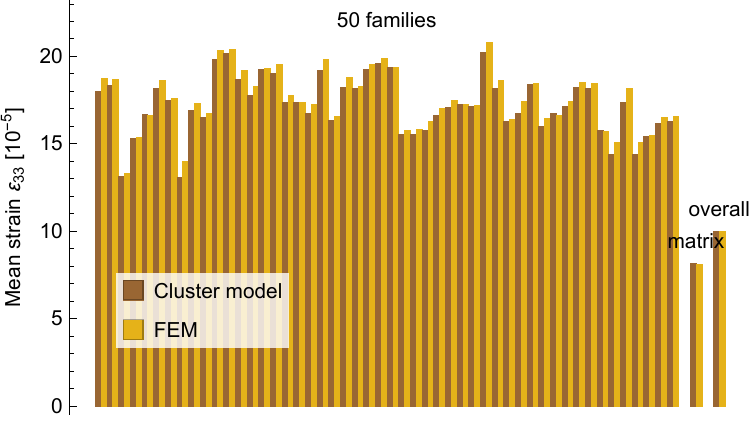}}
\hfill
\subfigure{(b)}{\includegraphics[angle=0,width=0.45\textwidth]{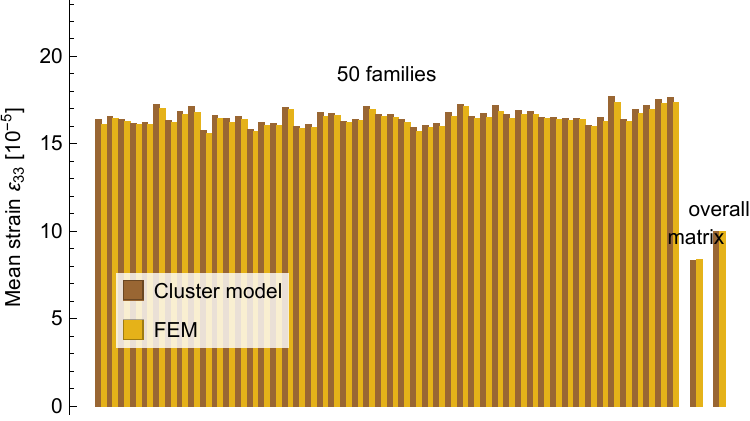}}
\caption{Mean strain $\varepsilon_{33}$ in all 50 voids, in the $\rm{AlSi_{12}}$ matrix and in the entire RUC, obtained by the cluster model and FEM for the microstructures in Fig.~\ref{Fig:example-rucs-c02}, (a) and (b) respectively, under uniaxial stress of the RUC along $x_3$ at global strain $\bar{\varepsilon}_{33}=10^{-4}$.}
\label{Fig:n50_t0c02_voids_bar_e33}
\end{figure}

\begin{figure}
\centering
\subfigure{(a)}{\includegraphics[angle=0,width=0.45\textwidth]{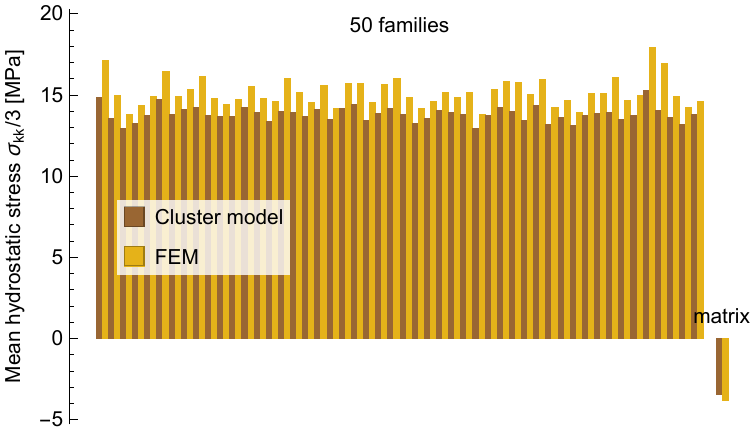}}
\hfill
\subfigure{(b)}{\includegraphics[angle=0,width=0.45\textwidth]{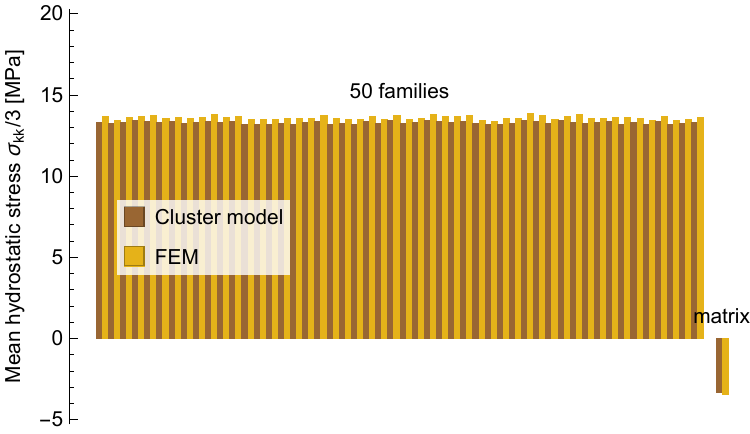}}
\caption{Mean hydrostatic stress $\sigma_{kk}/3$ in all 50 $\rm{SiC}$ inclusions and in the $\rm{AlSi_{12}}$ matrix, obtained by the cluster model and FEM for the microstructures in Fig.~\ref{Fig:example-rucs-c02}, (a) and (b) respectively, for unconstrained thermal expansion of the RUC under uniform temperature increase $\Delta T=10\degree\rm{C}$.}
\label{Fig:n50_t0c02_SiC_bar_skkT}
\end{figure}

A number of representative results of the performed calculations are presented in Figs.~\ref{Fig:KG20-Al2O3}--\ref{Fig:n50_t0c02_SiC_bar_skkT}, as well as in Tables~\ref{Tab:KGA-results-summary-Al2O3}--\ref{Tab:inclusions-results-summary-SiC-T}. In all tests, the discrepancy between the cluster model and FEM results for the effective bulk modulus, $\bar{K}$, and shear modulus, $\bar{G}$, was up to around $1\%$ and $4\%$, respectively, for the $\rm{Al_2O_3}$ and $\rm{SiC}$-based composites, an up to around $2.5\%$ and $2\%$, respectively, for the matrix with voids. The disrepancy for the effective thermal expansion coefficient, $\bar{\alpha}$, was below $1\%$. Overall, the mismatch between the cluster model and FEM was observed to increase with increasing volume fraction of inclusions/voids $f_{\ii}$ for all of the above-mentioned quantities. 

Example results for $f_{\ii}=20\%$ and all 30 random RUCs with different values of the packing parameter $\bar{\lambda}/(2R_{\ii})$ are shown in Figs.~\ref{Fig:KG20-Al2O3}--\ref{Fig:A20}. Fig.~\ref{Fig:KG20-Al2O3} shows the values of $\bar{K}$ (a) and $\bar{G}$ (b) for the $\rm{Al_2O_3}$-$\rm{AlSi_{12}}$ composite. It can be seen that the predicted values of both quantities decrease with increasing $\bar{\lambda}/(2R_{\ii})$, as is the difference between the cluster model and FEM results. Generally, the FEM results decrease with increasing $\bar{\lambda}/(2R_{\ii})$ faster than the cluster model results, displaying greater sensitivity to particle packing. It should be remarked that the $\rm{SiC}$-$\rm{AlSi_{12}}$ composite, results for which are not shown in a figure, behaves analogously to the $\rm{Al_2O_3}$-$\rm{AlSi_{12}}$ one. Similar results for $\rm{AlSi_{12}}$ with voids are shown in Fig.~\ref{Fig:KG20-voids}. Here, the values of $\bar{K}$ and $\bar{G}$ increase with increasing $\bar{\lambda}/(2R_{\ii})$, with the cluster model results being again less dependent on the packing parameter. Finally, in Fig.~\ref{Fig:A20}, one can see the estimates of $\bar{\alpha}$ for the $\rm{Al_2O_3}$-$\rm{AlSi_{12}}$ (a) and $\rm{SiC}$-$\rm{AlSi_{12}}$ (b) composites over the same packing-parameter range. The FEM results display yet again greater dependence on the packing parameter.

The results obtained for $f_{\ii}=10,30\%$, which are not presented in figures, behave analogously to those in Figs.~\ref{Fig:KG20-Al2O3}--\ref{Fig:A20} for $f_{\ii}=20\%$. One can observe that, in general, the estimates of $\bar{G}$ obtained by both methods are more scattered than those for $\bar{K}$ and $\bar{\alpha}$, especially towards higher values of $\bar{\lambda}/(2R_{\ii})$. This demonstrates a greater dependence of $\bar{G}$ and lesser dependence of $\bar{K}$ and $\bar{\lambda}$ on other spatial characteristics of composite microstructures besides the packing parameter. The nature of the dependence seems to be captured by the cluster model, in the sense that whenever there is an uptick or downtick in the FEM-computed value of $\bar{G}$ about its local mean, the cluster model estimate usually follows in the same direction.

Numerical values of the effective parameters $\bar{K}$, $\bar{G}$ and $\bar{\alpha}$ computed using the cluster model and FEM for two RUCs with extreme values of $\bar{\lambda}/(2R_{\ii})$ for each $f_{\ii}=10,20,30\%$ are collected in Table~\ref{Tab:KGA-results-summary-Al2O3} (for the $\rm{AlSi_{12}}-\rm{Al_2O_3}$ composite), Table~\ref{Tab:KGA-results-summary-SiC} (for the $\rm{AlSi_{12}}-\rm{SiC}$ composite) and in Table~\ref{Tab:KGA-results-summary-voids} (for $\rm{AlSi_{12}}$ with voids). In the tables, the Mori-Tanaka (MT) estimates are additionally provided for comparison. It can be seen that the MT values closely match the cluster model predictions obtained for the RUCs with the highest values of the packing parameter $\bar{\lambda}/(2R_{\ii})$---namely those in which inclusions/voids are more ``evenly'' distributed in the matrix. In other words, the cluster model results converge towards the MT estimates as $\bar{\lambda}/(2R_{\ii})$ increases.

\begin{table}
	\centering 
    {\small
	\begin{tabular}{cccccccc}
		\hline
		  $f_{\ii} [\%]$ & $\bar{\lambda}/(2R_{\ii})$ & $\bar{K}_C [\rm{GPa}]$ & $\bar{K}_{\rm{F}} [\rm{GPa}]$ & $\bar{G}_C [\rm{GPa}]$ & $\bar{G}_{\rm{F}} [\rm{GPa}]$ & $\bar{\alpha}_C [10^{-6}/\degree\rm{C}]$ & $\bar{\alpha}_{\rm{F}} [10^{-6}/\degree\rm{C}]$
		\\
		\hline
		  10 & 0.035 & 84.594 & 84.679 & 30.202 & 30.439 & 21.588 & 21.564 \\
		   & 0.815 & 84.558 & 84.580 & 30.151 & 30.277 & 21.598 & 21.592 \\
            & MT & 84.568 &  & 30.141 &  & 21.598 &  \\
		  20 & 0.008 & 92.319 & 92.574 & 35.195 & 35.820 & 19.571 & 19.512 \\
		   & 0.454 & 92.214 & 92.259 & 35.114 & 35.470 & 19.596 & 19.586 \\
           & MT & 92.209 &  & 35.039 &  & 19.597 &  \\
		  30 & 0.004 & 101.102 & 101.704 & 41.242 & 42.758 & 17.653 & 17.537 \\
		   & 0.280 & 100.923 & 101.044 & 40.863 & 41.812 & 17.688 & 17.664 \\
            & MT & 100.91 &  & 40.803 &  & 17.691 &  \\
		\hline
	\end{tabular}
    }
	\caption{Cluster model / FEM estimates of the effective bulk modulus, $\bar{K}_{\rm{C}}$ / $\bar{K}_{\rm{F}}$, shear modulus, $\bar{G}_{\rm{C}}$ / $\bar{G}_{\rm{F}}$, and thermal expansion coefficient, $\bar{\alpha}_{\rm{C}}$ / $\bar{\alpha}_{\rm{F}}$, for microstructures with the lowest and highest values of $\bar{\lambda}/(2R_{\ii})$ for $f_{\ii}=10,20,30\%$. As a reference, the estimates by the Mori-Tanaka (MT) model are provided. Composite: $\rm{Al_2O_3}$ inclusions in $\rm{AlSi_{12}}$ matrix.} 
	\label{Tab:KGA-results-summary-Al2O3}
\end{table}

\begin{table}
	\centering 
    {\small
	\begin{tabular}{cccccccc}
		\hline
		  $f_{\ii} [\%]$ & $\bar{\lambda}/(2R_{\ii})$ & $\bar{K}_C [\rm{GPa}]$ & $\bar{K}_{\rm{F}} [\rm{GPa}]$ & $\bar{G}_C [\rm{GPa}]$ & $\bar{G}_{\rm{F}} [\rm{GPa}]$ & $\bar{\alpha}_C [10^{-6}/\degree\rm{C}]$ & $\bar{\alpha}_{\rm{F}} [10^{-6}/\degree\rm{C}]$
		\\
		\hline
		  10 & 0.035 & 83.860 & 83.931 & 30.423 & 30.693 & 21.325 & 21.300 \\
		   & 0.815 & 83.830 & 83.846 & 30.365 & 30.502 & 21.336 & 21.330 \\
           & MT & 83.829 &  & 30.353 &  & 21.336 &  \\
		  20 & 0.008 & 90.661 & 90.877 & 35.720 & 36.455 & 19.047 & 18.983 \\
		   & 0.454 & 90.574 & 90.608 & 35.622 & 36.027 & 19.074 & 19.063 \\
             & MT & 90.570 &  & 35.538 &  & 19.075 &  \\
		  30 & 0.004 & 98.281 & 98.783 & 42.202 & 44.002 & 16.869 & 16.743 \\
		   & 0.280 & 98.135 & 98.228 & 41.764 & 42.862 & 16.908 & 16.883 \\
            & MT & 98.125 &  & 41.694 &  & 16.910 &  \\
		\hline
	\end{tabular}
    }
	\caption{Cluster model / FEM estimates of the effective bulk modulus, $\bar{K}_{\rm{C}}$ / $\bar{K}_{\rm{F}}$, shear modulus, $\bar{G}_{\rm{C}}$ / $\bar{G}_{\rm{F}}$, and thermal expansion coefficient, $\bar{\alpha}_{\rm{C}}$ / $\bar{\alpha}_{\rm{F}}$, for microstructures with the lowest and highest values of $\bar{\lambda}/(2R_{\ii})$ for $f_{\ii}=10,20,30\%$. As a reference, the estimates by the Mori-Tanaka (MT) model are provided. Composite: $\rm{SiC}$ inclusions in $\rm{AlSi_{12}}$ matrix.} 
	\label{Tab:KGA-results-summary-SiC}
\end{table}

\begin{table}
	\centering 
    {\small
	\begin{tabular}{cccccc}
		\hline
		  $f_{\ii} [\%]$ & $\bar{\lambda}/(2R_{\ii})$ & $\bar{K}_C [\rm{GPa}]$ & $\bar{K}_{\rm{F}} [\rm{GPa}]$ & $\bar{G}_C [\rm{GPa}]$ & $\bar{G}_{\rm{F}} [\rm{GPa}]$
		\\
		\hline
		  10 & 0.035 & 56.820 & 56.752 & 21.423 & 21.436 \\
		   & 0.815 & 57.143 & 57.386 & 21.493 & 21.597 \\
            & MT & 57.143 &  & 21.491 &  \\
		  20 & 0.008 & 42.290 & 41.809 & 17.491 & 17.343 \\
		   & 0.454 & 42.891 & 43.181 & 17.719 & 17.817 \\
           & MT & 42.912 &  & 17.705 &  \\
		  30 & 0.004 & 31.820 & 31.040 & 14.232 & 13.897 \\
		   & 0.280 & 32.467 & 32.757 & 14.388 & 14.379 \\
             & MT & 32.504 &  & 14.436 &  \\
		\hline
	\end{tabular}
    }
	\caption{Cluster model / FEM estimates of the effective bulk modulus, $\bar{K}_{\rm{C}}$ / $\bar{K}_{\rm{F}}$, and shear modulus, $\bar{G}_{\rm{C}}$ / $\bar{G}_{\rm{F}}$, for microstructures with the lowest and highest values of $\bar{\lambda}/(2R_{\ii})$ for $f_{\textup{i}}=10,20,30\%$. As a reference, the estimates by the Mori-Tanaka (MT) model are provided. Composite: $\rm{AlSi_{12}}$ matrix with voids.} 
	\label{Tab:KGA-results-summary-voids}
\end{table}

Tables~\ref{Tab:inclusions-results-summary-Al2O3}--\ref{Tab:inclusions-results-summary-SiC-T} and Figs.~\ref{Fig:n50_t0c02_Al2O3_bar_s33}--\ref{Fig:n50_t0c02_SiC_bar_skkT} present selected results regarding stresses/strains obtained by the cluster model and FEM in 50 individual inclusions/voids and in the matrix of two RUCs with extreme values of $\bar{\lambda}/(2R_{\ii})$ for $f_{\ii}=10,20,30\%$. Table~\ref{Tab:inclusions-results-summary-Al2O3} contains the results for the $\sigma_{33}$ stresses in the $\rm{Al_2O_3}$-$\rm{AlSi_{12}}$ composite under uniaxial stress in the $x_3$ direction at the global strain $\bar{\varepsilon}_{33}=10^{-4}$. It can be seen that while average stresses in the matrix predicted by both methods agree very well, the average values over all inclusions differ by up to $10\%$, with FEM predicting higher stresses. The added estimates by the Mori-Tanaka method agree quite well with the cluster model predictions. It can also be noticed that the standard deviations of mean values of $\sigma_{33}$ in all 50 inclusions, following from both the cluster model and FEM results, are much higher for the RUCs with small values of $\bar{\lambda}/(2R_{\ii})$ than for those with higher ones. This indicates, as can be expected, that the less evenly the inclusions are distributed in a composite, the greater is the variation of stresses among the inclusions. It also simultaneously shows that the cluster model is able to capture this effect. The computed mean values of $\sigma_{33}$ in all inclusions for $f_{\ii}=20\%$ are presented in Fig.~\ref{Fig:n50_t0c02_Al2O3_bar_s33}, where the difference in the stress variation between inclusions for RUCs with a small (a) and large (b) values of $\bar{\lambda}/(2R_{\ii})$ is clearly seen.

Table~\ref{Tab:inclusions-results-summary-SiC} provides analogous results, under uniaxial stress conditions, for stresses $\sigma_{33}$ in the $\rm{SiC}$-$\rm{AlSi_{12}}$ composite, which exhibit similar trends to the ones for $\rm{Al_2O_3}$-$\rm{AlSi_{12}}$ described above. Table~\ref{Tab:inclusions-results-summary-voids}, in turn, summarizes the results for the case of $\rm{AlSi_{12}}$ with voids. Here, the $\varepsilon_{33}$ strains are presented instead of stresses which vanish in the voids. Again, a good agreement between the cluster model and FEM results is observed for the matrix, except for $f_{\ii}=30\%$ and $\bar{\lambda}/(2R_{\ii})=0.004$ where the discrepancy reaches around $3\%$. For the voids, the relative difference in the mean values obtained by the cluster model and FEM is smaller than that for the stresses in composites with hard inclusions, reaching about $2.5\%$. As previously, the Mori-Tanaka estimates match closely the results for the RUCs with high values of the packing parameter. The dispersion of the mean values of $\varepsilon_{33}$ among the voids, as measured by the standard deviation, is much greater for the RUCs with lower values of $\bar{\lambda}/(2R_{\ii})$. This is also demonstrated more comprehensively in Fig.~\ref{Fig:n50_t0c02_voids_bar_e33} for the case of $f_{\ii}=20\%$.

The final two Tables~\ref{Tab:inclusions-results-summary-Al2O3-T} and \ref{Tab:inclusions-results-summary-SiC-T} list results obtained for the $\rm{Al_2O_3}$-$\rm{AlSi_{12}}$ and $\rm{SiC}$-$\rm{AlSi_{12}}$ composites, respectively, in the unconstrained thermal expansion tests under a uniform temperature increase $\Delta T=10\degree\rm{C}$. Since the tests themselves have an isotropic character, hydrostatic stresses $\sigma_{kk}/3$ in the matrix and inclusions were analyzed. The data obtained for both composites have similar features. There is a noticeable discrepancy between the cluster model and FEM results both for the inclusions and for the matrix, around $14\%$ in the case of $f_{\ii}=30\%$. Generally, as already noted before, the mismatch between the methods increases with increasing $f_{\ii}$ and decreasing $\bar{\lambda}/(2R_{\ii})$, meaning that denser and more closely-packed composites are more difficult to accurately model using the cluster scheme, at least in the hard-inclusion case. The cluster model can be also observed to consistently underestimate the hydrostatic stresses, in terms of their absolute values, with respect to the FEM results. The greater dispersion of mean values in individual inclusions for the RUCs with lower values of $\bar{\lambda}/(2R_{\ii})$ is yet again apparent in Tables~\ref{Tab:inclusions-results-summary-Al2O3-T}--\ref{Tab:inclusions-results-summary-SiC-T} and in Fig.~\ref{Fig:n50_t0c02_SiC_bar_skkT} for the case of $\rm{SiC}$ inclusions at the $f_{\ii}=20\%$ volume fraction.

The last remark about the results for individual inclusions is that the distributions of mean stresses or strains in all inclusions/voids obtained by the cluster model and FEM have similar shapes, as can be seen in Figs.~\ref{Fig:n50_t0c02_Al2O3_bar_s33}--\ref{Fig:n50_t0c02_SiC_bar_skkT}. In other words, in some respects similarly to the case of effective values of $\bar{G}$ discussed earlier, if the mean stress or strain in one inclusion is greater than in another one according to FEM, then it is usually also greater according to the cluster model. This shows that the cluster model correctly reflects the variability of the mean stresses or strains in individual inclusions/voids.

\begin{table}
	\centering 
    {\small
	\begin{tabular}{cccccccc}
		\hline
		  $f_{\ii} [\%]$ & $\bar{\lambda}/(2R_{\ii})$ & $\sigma_{33}^{\rm{iC}} [\rm{MPa}]$  & $\hat{\sigma}_{33}^{\rm{iC}} [\rm{MPa}]$ & $\sigma_{33}^{\rm{iF}} [\rm{MPa}]$ & $\hat{\sigma}_{33}^{\rm{iF}} [\rm{MPa}]$ & $\sigma_{33}^{\rm{mC}} [\rm{MPa}]$ & $\sigma_{33}^{\rm{mF}} [\rm{MPa}]$
		\\
		\hline
		  10 & 0.035 & 12.815 & 1.240 & 13.461 & 1.583 & 7.603 & 7.602 \\ 
          & 0.815 & 12.386 & 0.161 & 12.686 & 0.170 & 7.610 & 7.611 \\ 
          & MT & 12.332 &  &  &  & 7.610 &  \\ 
        20 & 0.008 & 13.487 & 1.140 & 14.244 & 1.470 & 8.311 & 8.304 \\ 
          & 0.454 & 13.504 & 0.288 & 13.934 & 0.333 & 8.310 & 8.310 \\ 
           & MT & 13.414 &  &  &  & 8.310 &  \\ 
        30 & 0.004 & 15.148 & 1.351 & 16.509 & 1.815 & 9.095 & 9.052 \\ 
          & 0.280 & 14.757 & 0.451 & 15.517 & 0.553 & 9.122 & 9.116 \\ 
           & MT & 14.673 &  &  &  & 9.236 &  \\ 
		\hline
	\end{tabular}
    }
	\caption{Cluster model / FEM estimates of the $\sigma_{33}$ stress: mean in all inclusions, $\sigma_{33}^{\rm{iC}}$ / $\sigma_{33}^{\rm{iF}}$, standard deviation of means in individual inclusions: $\hat{\sigma}_{33}^{\rm{iC}}$ / $\hat{\sigma}_{33}^{\rm{iF}}$, and mean in the matrix: $\sigma_{33}^{\rm{mC}}$ / $\sigma_{33}^{\rm{mF}}$, for microstructures with the lowest and highest values of $\bar{\lambda}/(2R_{\ii})$ for $f_{\ii}=10,20,30\%$. Composite: $\rm{Al_2O_3}$ inclusions in $\rm{AlSi_{12}}$ matrix. Uniaxial stress along $x_3$ at global strain $\bar{\varepsilon}_{33}=10^{-4}$.} 
	\label{Tab:inclusions-results-summary-Al2O3}
\end{table}

\begin{table}
	\centering 
    {\small
	\begin{tabular}{cccccccc}
		\hline
		  $f_{\ii} [\%]$ & $\bar{\lambda}/(2R_{\ii})$ & $\sigma_{33}^{\rm{iC}} [\rm{MPa}]$  & $\hat{\sigma}_{33}^{\rm{iC}} [\rm{MPa}]$ & $\sigma_{33}^{\rm{iF}} [\rm{MPa}]$ & $\hat{\sigma}_{33}^{\rm{iF}} [\rm{MPa}]$ & $\sigma_{33}^{\rm{mC}} [\rm{MPa}]$ & $\sigma_{33}^{\rm{mF}} [\rm{MPa}]$
		\\
		\hline
		  10 & 0.035 & 13.012 & 1.306 & 13.722 & 1.691 & 7.634 & 7.635 \\ 
          & 0.815 & 12.555 & 0.170 & 12.876 & 0.180 & 7.641 & 7.642 \\ 
           & MT & 12.496 &  &  &  & 7.634 &  \\ 
        20 & 0.008 & 13.724 & 1.197 & 14.552 & 1.553 & 8.379 & 8.381 \\ 
          & 0.454 & 13.735 & 0.304 & 14.204 & 0.354 & 8.378 & 8.381 \\ 
           & MT & 13.637 &  &  &  & 8.376 &  \\ 
        30 & 0.004 & 15.488 & 1.423 & 17.025 & 1.942 & 9.212 & 9.182 \\ 
          & 0.280 & 15.064 & 0.479 & 15.909 & 0.594 & 9.238 & 9.240 \\  
           & MT & 14.971 &  &  &  & 9.236 &  \\ 
		\hline
	\end{tabular}
    }
	\caption{Cluster model / FEM estimates of the $\sigma_{33}$ stress: mean in all inclusions, $\sigma_{33}^{\rm{iC}}$ / $\sigma_{33}^{\rm{iF}}$, standard deviation of means in individual inclusions: $\hat{\sigma}_{33}^{\rm{iC}}$ / $\hat{\sigma}_{33}^{\rm{iF}}$, and mean in the matrix: $\sigma_{33}^{\rm{mC}}$ / $\sigma_{33}^{\rm{mF}}$, for microstructures with the lowest and highest values of $\bar{\lambda}/(2R_{\ii})$ for $f_{\ii}=10,20,30\%$. Composite: $\rm{SiC}$ inclusions in $\rm{AlSi_{12}}$ matrix. Uniaxial stress along $x_3$ at global strain $\bar{\varepsilon}_{33}=10^{-4}$.} 
	\label{Tab:inclusions-results-summary-SiC}
\end{table}

\begin{table}
	\centering 
    {\small
	\begin{tabular}{cccccccc}
		\hline
		  $f_{\ii} [\%]$ & $\bar{\lambda}/(2R_{\ii})$ & $\varepsilon_{33}^{\rm{iC}} [10^{-4}]$  & $\hat{\varepsilon}_{33}^{\rm{iC}} [10^{-4}]$ & $\varepsilon_{33}^{\rm{iF}} [10^{-4}]$ & $\hat{\varepsilon}_{33}^{\rm{iF}} [10^{-4}]$ & $\varepsilon_{33}^{\rm{mC}} [10^{-4}]$ & $\varepsilon_{33}^{\rm{mF}} [10^{-4}]$
		\\
		\hline
		  10 & 0.035 & 1.7944 & 0.1861 & 1.7952 & 0.1730 & 0.9117 & 0.9117 \\ 
          & 0.815 & 1.8098 & 0.0323 & 1.7728 & 0.0302 & 0.9100 & 0.9142 \\ 
           & MT & 1.8150 &  &  &  & 0.9094 &  \\ 
        20 & 0.008 & 1.7230 & 0.1666 & 1.7556 & 0.1667 & 0.8192 & 0.8113 \\ 
          & 0.454 & 1.6618 & 0.0458 & 1.6448 & 0.0423 & 0.8345 & 0.8390 \\ 
           & MT & 1.6647 &  &  &  & 0.8338 &  \\ 
        30 & 0.004 & 1.5578 & 0.1922 & 1.6062 & 0.2002 & 0.7609 & 0.7405 \\ 
          & 0.280 & 1.5404 & 0.0601 & 1.5404 & 0.0550 & 0.7684 & 0.7687 \\   
           & MT & 1.5370 &  &  &  & 0.7699 &  \\ 
		\hline
	\end{tabular}
    }
	\caption{Cluster model / FEM estimates of the $\varepsilon_{33}$ strain: mean in all inclusions, $\varepsilon_{33}^{\rm{iC}}$ / $\varepsilon_{33}^{\rm{iF}}$, standard deviation of means in individual inclusions: $\hat{\varepsilon}_{33}^{\rm{iC}}$ / $\hat{\varepsilon}_{33}^{\rm{iF}}$, and mean in the matrix: $\varepsilon_{33}^{\rm{mC}}$ / $\varepsilon_{33}^{\rm{mF}}$, for microstructures with the lowest and highest values of $\bar{\lambda}/(2R_{\ii})$ for $f_{\ii}=10,20,30\%$. Composite: $\rm{AlSi_{12}}$ matrix with voids. Uniaxial stress along $x_3$ at global strain $\bar{\varepsilon}_{33}=10^{-4}$.} 
	\label{Tab:inclusions-results-summary-voids}
\end{table}

\begin{table}
	\centering 
    {\small
	\begin{tabular}{cccccccc}
		\hline
		  $f_{\ii} [\%]$ & $\bar{\lambda}/(2R_{\ii})$ & $\sigma_{\rm{h}}^{\rm{iC}} [\rm{MPa}]$  & $\hat{\sigma}_{\rm{h}}^{\rm{iC}} [\rm{MPa}]$ & $\sigma_{\rm{h}}^{\rm{iF}} [\rm{MPa}]$ & $\hat{\sigma}_{\rm{h}}^{\rm{iF}} [\rm{MPa}]$ & $\sigma_{\rm{h}}^{\rm{mC}} [\rm{MPa}]$ & $\sigma_{\rm{h}}^{\rm{mF}} [\rm{MPa}]$
		\\
		\hline
		  10 & 0.035 & 13.946 & 0.224 & 14.880 & 0.370 & -1.550 & -1.650 \\ 
          & 0.815 & 13.587 & 0.017 & 13.882 & 0.063 & -1.510 & -1.540 \\ 
            & MT & 13.582 &  &  &  & -1.509 &  \\ 
        20 & 0.008 & 12.247 & 0.393 & 13.377 & 0.717 & -3.062 & -3.336 \\ 
          & 0.454 & 11.803 & 0.067 & 12.079 & 0.094 & -2.951 & -3.014 \\ 
            & MT & 11.782 &  &  &  & -2.945 &  \\ 
        30 & 0.004 & 10.514 & 0.524 & 11.960 & 0.969 & -4.506 & -5.113 \\ 
          & 0.280 & 10.096 & 0.107 & 10.463 & 0.123 & -4.327 & -4.474 \\ 
            & MT & 10.067 &  &  &  & -4.314 &  \\ 
		\hline
	\end{tabular}
    }
	\caption{Cluster model / FEM estimates of the hydrostatic stress $\sigma_{\rm{h}}=\sigma_{kk}/3$: mean in all inclusions, $\sigma_{\rm{h}}^{\rm{iC}}$ / $\sigma_{\rm{h}}^{\rm{iF}}$, standard deviation of means in individual inclusions: $\hat{\sigma}_{\rm{h}}^{\rm{iC}}$ / $\hat{\sigma}_{\rm{h}}^{\rm{iF}}$, and mean in the matrix: $\sigma_{\rm{h}}^{\rm{mC}}$ / $\sigma_{\rm{h}}^{\rm{mF}}$, for microstructures with the lowest and highest values of $\bar{\lambda}/(2R_{\ii})$ for $f_{\ii}=10,20,30\%$. Composite: $\rm{Al_2O_3}$ inclusions in $\rm{AlSi_{12}}$ matrix. Unconstrained thermal expansion under a uniform temperature increase $\Delta T=10\degree\rm{C}$.} 
	\label{Tab:inclusions-results-summary-Al2O3-T}
\end{table}

\begin{table}
	\centering 
    {\small
	\begin{tabular}{cccccccc}
		\hline
		  $f_{\ii} [\%]$ & $\bar{\lambda}/(2R_{\ii})$ & $\sigma_{\rm{h}}^{\rm{iC}} [\rm{MPa}]$  & $\hat{\sigma}_{\rm{h}}^{\rm{iC}} [\rm{MPa}]$ & $\sigma_{\rm{h}}^{\rm{iF}} [\rm{MPa}]$ & $\hat{\sigma}_{\rm{h}}^{\rm{iF}} [\rm{MPa}]$ & $\sigma_{\rm{h}}^{\rm{mC}} [\rm{MPa}]$ & $\sigma_{\rm{h}}^{\rm{mF}} [\rm{MPa}]$
		\\
		\hline
		  10 & 0.035 & 15.698 & 0.261 & 16.785 & 0.435 & -1.744 & -1.862 \\ 
          & 0.815 & 15.280 & 0.020 & 15.608 & 0.069 & -1.698 & -1.731 \\ 
            & MT & 15.274 &  &  &  & -1.697 &  \\ 
        20 & 0.008 & 13.826 & 0.457 & 15.168 & 0.849 & -3.457 & -3.781 \\ 
          & 0.454 & 13.307 & 0.077 & 13.615 & 0.107 & -3.327 & -3.397 \\ 
            & MT & 13.282 &  &  &  & -3.320 &  \\ 
        30 & 0.004 & 11.901 & 0.612 & 13.625 & 1.161 & -5.101 & -5.820 \\ 
          & 0.280 & 11.408 & 0.125 & 11.826 & 0.143 & -4.889 & -5.056 \\  
            & MT & 11.374 &  &  &  & -4.875 &  \\ 
		\hline
	\end{tabular}
    }
	\caption{Cluster model / FEM estimates of the hydrostatic stress $\sigma_{\rm{h}}=\sigma_{kk}/3$: mean in all inclusions, $\sigma_{\rm{h}}^{\rm{iC}}$ / $\sigma_{\rm{h}}^{\rm{iF}}$, standard deviation of means in individual inclusions: $\hat{\sigma}_{\rm{h}}^{\rm{iC}}$ / $\hat{\sigma}_{\rm{h}}^{\rm{iF}}$, and mean in the matrix: $\sigma_{\rm{h}}^{\rm{mC}}$ / $\sigma_{\rm{h}}^{\rm{mF}}$, for microstructures with the lowest and highest values of $\bar{\lambda}/(2R_{\ii})$ for $f_{\ii}=10,20,30\%$. Composite: $\rm{SiC}$ inclusions in $\rm{AlSi_{12}}$ matrix. Unconstrained thermal expansion under a uniform temperature increase $\Delta T=10\degree\rm{C}$.} 
	\label{Tab:inclusions-results-summary-SiC-T}
\end{table}

\section{Conclusion}
In the paper, efficient computational procedures for the multi-family variant of the cluster model were proposed.  A broad verification of the method for two composites and a porous matrix with random inhomogeneity distribution has been performed by comparing the results of the analytical model with outcomes of numerical homogenization. To our best knowledge, a cluster model has not been yet applied and verified in the present context.  

Analysis of the effect of packing on the overall thermoelastic properties and local response of particulate composites has been performed. Very good agreement between the mean-field model predictions and FEM results, both concerning the overall properties and estimates of local mean strains and stresses for individual inclusions, has been demonstrated. In agreement with previous experimental and numerical studies we observed that while clustering has minimal influence on the overall elastic properties, it leads to a greater dispersion of mean fields between particles. The overall elastic stiffness increased only slightly with decreasing mean nearest-neighbour distance between inclusions, while at the same time such a change in the microstructure substantially affected the standard deviation of mean stresses for individual inclusions in the representative unit cells. As far as we are aware, a mean-field model for thermoelastic metal matrix composites with comparable computational efficiency and predictive capabilities regarding the response of individual inclusions in the representative volume is not available in the existing literature.

The proposed approach is planned to be used in studying damage development in composite materials as it enables assessment of different thresholds of damage initiation depending on the spatial distribution of randomly-placed particles in a representative composite volume. In the present research, the particles were assumed to be spherical and of equal size, while the considered examples differed in the volume fraction of inclusions and their mean nearest-neighbour distances. In the future, microstructures with different inhomogeneity sizes or ellipsoidal particle shapes are intended to be studied. 

\section*{Acknowledgements}
K.K.-G. acknowledges the financial support provided to the research by the European Union’s Research and Innovation Program Horizon Europe under the Grant Agreement No. 101086342 – HORIZON-MSCA-2021-SE-01 Project DIAGONAL (Ductility and fracture toughness analysis of Functionally Graded Materials), co-founded by the Polish Ministry of Science and Higher Education under the program ''International Projects Co-financed''. P.H. and M. M. acknowledge the support of the National Science Centre through the {project} 2021/41/B/ST8/03345 within which procedures for computational homogenization of representative unit cells with random distribution of inhomogeneities have been developed.

\appendix
\section{Computational algorithm for the multi-family variant of the cluster model\label{Sec:App1}}

In order to find the effective properties \eqref{Eq:C} and \eqref{Eq:Beta} of the composite with the use of the cluster interaction model, the averaged interaction tensors \eqref{Eq:AGamma} need to be calculated for each pair $I$ and $K$ of families. These tensors are required to specify the localization tensors $\mathbb{A}_K$.

The computational algorithm, which is followed to compute effective thermoelastic properties and localization tensors for the interaction cluster model in the case of a particulate two-phase composite with spherical inclusions, is composed of the following steps (isotropic properties of the matrix phase are assumed):

\begin{enumerate}
\item[\textbf{Step 1:}] {Input data are read}, including 
\begin{itemize}
\item[--] local properties of the two phases: $\mathbb{C}_r$, $\boldsymbol{\beta}_r$ ($r=\rm{m},\rm{i}$),
\item[--] the set of locations $(x_I,y_I,z_I)$ and radii $r_I$  of particles ($I=1,\ldots,N$) in the elementary cubic volume $1 \times 1 \times 1$ of the composite,
\item[--] the cluster size $R_c$. 
\end{itemize}
\item[\textbf{Step 2:}] The averaged interaction tensors \eqref{Eq:AGamma}  are calculated for each pair $(I,K)$ of inclusion families. 
Details of the algorithm for finding the tensor 
$\bar{\boldsymbol{\Gamma}}^{IK}$ for a given pair of families are discussed below. The polarisation tensor $\mathbb{P}_0$ is calculated for the isotropic matrix material by the formula 
\begin{equation}\label{Eq:Po}
\mathbb{P}_0=\frac{1-2\nu_{\mm}}{6\mu_{\mm}(1-\nu_{\mm})}\mathbb{I}_P+\frac{2(4-5\nu_{\mm})}{15\mu_{\mm}}\mathbb{I}_D\,,
\end{equation}
where $\mu_{\mm}$ and $\nu_{\mm}$ are the shear modulus and Poisson's ratio of the isotropic matrix material, and $\mathbb{I}_P$ and $\mathbb{I}_D$ are orthogonal projectors to the subspaces of hydrostatic and deviatoric second-order tensors, respectively.
\item[\textbf{Step 3:}] The fourth-order tensors $\mathbb{M}^{IK}$ and the second-order tensors $\mathbf{w}^K$ ($I,K=1,\ldots,N$) are calculated using the formulas \eqref{Eq:MIK} and \eqref{Eq:wI}.
\item[\textbf{Step 4:}] The sets of fourth-order tensorial equations \eqref{Eq:set-thermoel} and second-order tensorial equations \eqref{Eq:set-thermoel2} are solved using the extended Gauss elimination procedure discussed below. By solving these sets of equations, the fourth-order localization tensors $\mathbb{A}_I$ and the second-order tensors $\mathbf{e}_I$ are found. Next, $\mathbf{A}_{\mm}$ and $\mathbf{e}_{\mm}$ are found using \eqref{Eq:Am} and \eqref{Eq:em}.
\item[\textbf{Step 5:}] The effective elastic stiffness tensor $\bar{\mathbb{C}}$ and thermal stress tensor $\bar{\boldsymbol{\beta}}$ are calculated using relations \eqref{Eq:C} and \eqref{Eq:Beta}. 
\end{enumerate}
It should be noted that despite the assumed random distribution of particles the effective properties are not perfectly isotropic due to the finite number of particles in the representative unit cell.  

\noindent\textbf{The computationally efficient procedure for finding the $\bar{\boldsymbol{\Gamma}}^{IK}$ tensor} for a given pair $(I,K)$ of families in step 2 of the algorithm above proceeds as follows:
\begin{enumerate}
\item[\textbf{Step 2.1:}] {Input data are read}, including 
\begin{itemize}
\item[--] elastic constants of the isotropic matrix: $\mu_{\mm}$, $\nu_{\mm}$,
\item[--] locations $(x_I,y_I,z_I)$ and $(x_K,y_K,z_K)$ as well as radii $r_I$ and $r_K$  of two particles in the elementary unit volume $1 \times 1 \times 1$ of the composite,
\item[--] the cluster size $R_c$\,.
\end{itemize}
\item[\textbf{Step 2.2:}] If $I=K$, by translating the inclusion $I$ to the frame origin and repeating it periodically in each of the three directions, the regular cubic (RC) arrangement of particles is obtained. Using this observation, the closed form relations derived in \cite{Bieniek24} are used to compute the $\bar{\boldsymbol{\Gamma}}^{II}$ tensor.
\item[\textbf{Step 2.3:}] If $I<K$, 
\item [--] the pair of inclusions is translated in space so that inclusion $I$ is placed at the frame origin: $(x^*_I,y^*_I,z^*_I)=(0,0,0)$, while inclusion $K$ is moved to the position $(x^*_K,y^*_K,z^*_K)=(x_K-x_I,y_K-y_I,z_K-z_I)$, 
\item[--] inclusion $K$ is periodically repeated in space in three directions to form an infinite set of inclusions $(x^{*(K)}_j,y^{*(K)}_j,z^{*(K)}_j)$ ($j=1,\ldots,\infty$) belonging to the family $K$,
\item[--] to form a cluster, inclusions $j=1,\ldots, N_c$ of the family $K$ which satisfy the condition $(x^{*(K)}_j)^2+(y^{*(K)}_j)^2+(z^{*(K)}_j)^2\leq R_c^2$ are selected,
\item[--] the tensor $\bar{\boldsymbol{\Gamma}}^{IK}$ is calculated by summing the tensors ${\boldsymbol{\Gamma}}^{Ij}$. The latter are calculated using the relations (B.1-B.6) in \cite{Bieniek24} in which $R$ is the distance between inclusion $I$ and $j$-th inclusion of the family $K$, so that $R=\sqrt{(x^{*(j)}_K)^2+(y^{*(j)}_K)^2+(z^{*(j)}_K)^2}$, components of the unit vector $\mathbf{n}$ are $n_k=(x_K^{*(j)}/R,y_K^{*(j)}/R,z_K^{*(j)}/R)$, while $a$ and $b$ are radii $r_I$ and $r_K$, respectively.
\item[\textbf{Step 2.4:}] If $I>K$, the symmetry property of the tensor $\bar{\boldsymbol{\Gamma}}^{IK}$ is used \cite{Berveiller87}, namely (no summation over $I$ and $K$)  $$r_I^3\bar{\boldsymbol{\Gamma}}^{IK}=r_K^3\bar{\boldsymbol{\Gamma}}^{KI}\,.$$
\end{enumerate}

\noindent\textbf{The extended Gauss elimination procedure} solves the set of the tensorial equations of the form 
$[\mathbf{M}].[\mathbf{X}] = [\mathbf{b}]$, where $[\mathbf{M}]$ is a known $N \times N$ matrix of second-order tensors $\mathbf{M}^{IK}$ ($I,K=1,\ldots,N$) in $E^k\otimes E^k$ tensorial space, 
$[\mathbf{X}]$ and $[\mathbf{b}]$ are column vectors of $N$ unknown and known tensors $\mathbf{X}_K$ and $\mathbf{b}^I$, respectively, in $E^k\otimes E^l$ tensorial space. When applied to equation set \eqref{Eq:set-thermoel}, $k=l=6$, as the fourth-order tensors $\mathbb{M}^{IK}$, $\mathbb{A}_K$ and $\mathbb{I}$ are written as second-order tensors in six-dimensional space (see \cite{Kowalczyk09b}). When applied to \eqref{Eq:set-thermoel2}, $k=6$ while $l=1$, as the second-order tensors $\mathbf{e}^{K}$ and $\mathbf{w}_I$ are written as vectors in six-dimensional space. The first stage of the algorithm is aimed to transform the original set of equations into $[\mathbf{M}_{*}].[\mathbf{X}] = [\mathbf{b}_{*}]$, in which in the matrix $[\mathbf{M}_{*}]$ the diagonal tensorial components are identity tensors: $\mathbf{M}_{*}^{II}=\mathbf{I}$, while components $\mathbf{M}_{*}^{IK}$ for which $I>K$ are equal to $\mathbf{0}$. Next, the solution for $[\mathbf{X}]$ is found by substitution, starting from the element $\mathbf{X}_{N}=\mathbf{b}_{*}^{N}$. The detailed algorithm proceeds as follows:
\begin{enumerate}
\item The sought variables $[\mathbf{M}_{*}]$ and $[\mathbf{b}_{*}]$ are preset to $[\mathbf{0}]$, while iterative variables $[\mathbf{M}_{\rm{trial}}]$ and $[\mathbf{b}_{\rm{trial}}]$ are preset to $[\mathbf{M}]$ and $[\mathbf{b}]$. 
\item For $I=1,\ldots,N-1$ we perform the steps:
\begin{itemize}
\item calculate (no summation over $I$) $$\mathbf{b}^{I}_{*}=(\mathbf{M}^{II}_{\rm{trial}})^{-1}\mathbf{b}^{I}_{\rm{trial}}$$
    \item for $K=I,\ldots, N$ calculate
    $$\mathbf{M}^{IK}_{*}=(\mathbf{M}^{II}_{\rm{trial}})^{-1}(\mathbf{M}^{IK}_{\rm{trial}})$$
    \item replace $\mathbf{M}_{\rm{trial}}^{IK}$ and $\mathbf{b}_{\rm{trial}}^I$ with $\mathbf{M}_{*}^I$ and $\mathbf{b}_{*}^I$, respectively\,.
  \item For $K=I+1,\ldots,N$ calculate
   \begin{itemize}
      \item for  $L=I+1,\ldots,N$ 
      calculate $$\mathbf{M}^{KL}_{*}=\mathbf{M}^{KL}_{\rm{trial}}-\mathbf{M}^{KI}_{\rm{trial}}\mathbf{M}^{IL}_{\rm{trial}}$$ and replace
      $\mathbf{M}_{\rm{trial}}^{KL}=\mathbf{M}_{*}^{KL}$,
      \item calculate $$\mathbf{b}_{*}^K=\mathbf{b}_{\rm{trial}}^K-\mathbf{M}_{\rm{trial}}^{KI}\mathbf{b}^I_{\rm{trial}}$$ and replace $\mathbf{b}_{\rm{trial}}^K=\mathbf{b}_{*}^K$.
   \end{itemize}
\end{itemize}   
\item Calculate $\mathbf{b}_{*}^{N}=(\mathbf{M}_{\rm{trial}}^{NN})^{-1}\mathbf{b}^{N}_{\rm{trial}}$.
\item Preset $[\mathbf{X}]=[\mathbf{0}]$ 
\item Set $\mathbf{X}_{N}=\mathbf{b}_{*}^{N}$
\item For $I=1,\ldots, N-1$ calculate
$$\mathbf{X}_{N-I}=\mathbf{b}_{*}^{N-I}+\sum_{K=N-I+1}^{N}\mathbf{M}^{N-I,K}_{*}\mathbf{X}_K$$
 
\end{enumerate}
 
Having found the effective thermoelastic properties, assuming some macroscopic stress tensor $\bar{\boldsymbol{\sigma}}$ or strain tensor $\bar{\boldsymbol{\varepsilon}}$ and temperature increase $\theta$, the respective local quantities can be found. For example, mean strains and stresses in all particles in the representative unit cell for the uniaxial tension process in direction 1 up to tensile strain $\varepsilon_{11}=0.01$, with no temperature change ($\theta=0$), can be found assuming:
\begin{displaymath}
 \bar{\boldsymbol{\sigma}}\sim\left[\begin{array}{ccc}
 \Sigma& 0 & 0 \\ 0 & 0 & 0 \\ 0 & 0 & 0
 \end{array}\right]   \quad 
  \bar{\boldsymbol{\varepsilon}}\sim\left[\begin{array}{ccc}
 0.01 & E_{12} & E_{13} \\ E_{12} & E_{22} & E_{23} \\ E_{13} & E_{23} & E_{33}
 \end{array}\right] 
\end{displaymath}
Next, using the overall relation \eqref{Eq:tot} in the reduced form $\bar{\boldsymbol{\sigma}}=\bar{\mathbb{C}}\cdot  \bar{\boldsymbol{\varepsilon}}$,  we find the six unknowns: $\Sigma, E_{22}, E_{33},E_{12}, E_{13},E_{23}$. After finding the overall strain $\bar{\boldsymbol{\varepsilon}}$, mean strains and stresses for each inclusion in the RUC are found by the localization relation and local constitutive law: $\boldsymbol{\varepsilon}_I=\mathbb{A}_I\cdot\bar{\boldsymbol{\varepsilon}}$, $\boldsymbol{\sigma}_I=\mathbb{C}_{\ii}\cdot\boldsymbol{\varepsilon}_I$. 
If one considers a cooling or heating process with a temperature change $\theta$ under zero external stress, thermal strains and stresses in each inclusion and the matrix can be found assuming $\bar{\boldsymbol{\sigma}}=\textbf{0}$ and solving first the overall relation
$\bar{\mathbf{C}}\cdot\bar{\boldsymbol{\varepsilon}}-\bar{\boldsymbol{\beta}}\theta=\mathbf{0}$ for $\bar{\boldsymbol{\varepsilon}}$. Next, applying the localization relations \eqref{Eq:loc} and the constitutive laws \eqref{Eq:tel}, one finds mean strains and stresses for each inclusion in the RUC and the matrix. Alternatively, in both cases, mean stresses in inclusions can be found by using stress localization tensors $\mathbb{B}_I$ and $\mathbf{s}_I$,
\begin{displaymath}
\mathbb{B}_I=\mathbb{C}_I\mathbb{A}_I\bar{\mathbb{C}}^{-1}\Rightarrow \boldsymbol{\sigma}_I=\mathbb{B}_I\cdot\bar{\boldsymbol{\sigma}}+\mathbf{s}_I\theta\,.
\end{displaymath}

\section{Convergence of the cluster model\label{Sec:App2}}

\begin{figure}
\centering
\subfigure{(a)}{\includegraphics[angle=0,width=0.45\textwidth]{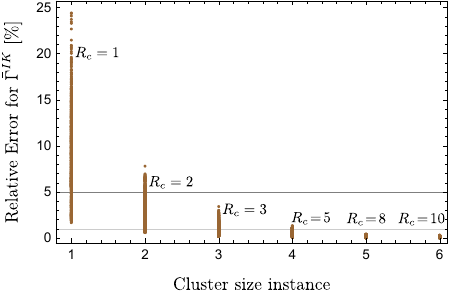}}
\hfill
\subfigure{(b)}{\includegraphics[angle=0,width=0.45\textwidth]{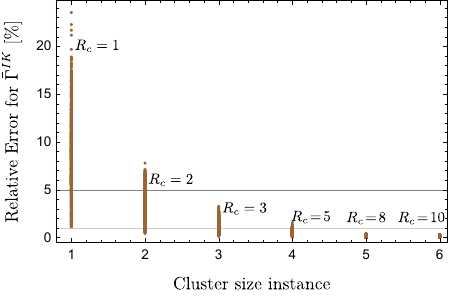}}
\caption{Relative errors \eqref{Eq:error} for all the components of all $\bar{\boldsymbol{\Gamma}}^{KL}$, ($K,L=1,\ldots, 50$) for two selected microstructure realizations with (a) high ($\bar{\lambda}/(2R_{\ii})=0.28$) and (b) low ($\bar{\lambda}/(2R_{\ii})=0.0043$) mean minimum distance between the inclusions with a total volume fraction of $30\%$.}
\label{Fig:error}
\end{figure}

As discussed above, the interaction cluster scheme requires one to define a cluster size $R_c$ -- a size of the domain for which direct interactions between a given pair of inclusions are accounted for (see Fig. \ref{Fig:clusterradius}). To establish a value which is large enough to obtain reliable results, we have calculated the averaged interaction tensors $\bar{\boldsymbol{\Gamma}}^{IK}$ for selected microstructure realizations increasing the cluster size from $R_c=1$ to $R_c=20$, assuming the representative unit volume to be a $1\times 1\times 1$ cube, and we have analyzed how the components $\Gamma_{ijkl}^{KL}$ of these tensors change with an increasing cluster size. The relative error for a given component is defined as the following quantity
\begin{equation}\label{Eq:error}
    \varepsilon_{rel}=\frac{|\Gamma_{ijkl}^{KL}(R_c=20)-\Gamma_{ijkl}^{KL}(R_c)|}{\Gamma_{ijkl}^{KL}(R_c=20)}\times 100\%\end{equation}
The results of the study for two microstructure realizations with a low and high mean minimum distance and the inclusion volume fraction of $30\%$ are demonstrated in Fig. \ref{Fig:error}. It is seen that for the cluster size $R_c=3$ the relative error drops below $5\%$, while for $R_c=8$ it drops below $1\%$. Based on this outcome, in the calculations reported in this paper the cluster size $R_c=10$ was assumed. 

\bibliography{references,num-references}

\end{document}